\documentclass[letterpaper,titlepage,12pt,dvipsnames]{article}\pdfoutput=1
\usepackage[english]{babel}
\usepackage{amsmath,amssymb,amsthm,mathrsfs,bbm}
\usepackage{latexsym,amscd,amsbsy,amsfonts,dsfont}
\usepackage{graphicx}
\usepackage{physics}
\usepackage{cancel}
\usepackage{comment}
\usepackage[
      colorlinks=true,
      linktocpage=true,
      breaklinks=true,
      linkcolor=blue,
      urlcolor=Purple,
      filecolor=black,
      citecolor=purple,
      menucolor=black,
      pdfstartview=FitV,
      bookmarksopen=true
      ]{hyperref}
\usepackage{tikz}

\usepackage{xcolor}

\usepackage[normalem]{ulem}
\definecolor{forestgreen}{rgb}{0.13, 0.54, 0.13}

\numberwithin{equation}{section}

\begin{document}

\begin{titlepage}

\leftline{}

\centerline{\Large \bf Semiclassical effects on ultramassive spacetimes} 
\bigskip

\vskip 0.5cm
\centerline{\large \bf Valentin Boyanov$^{a,b}$, Luis J. Garay$^{c}$}
\smallskip
\centerline{\large \bf Alberto Garc\'ia Mart\'in-Caro$^{d,e}$, Gerardo Garc\'ia-Moreno$^{f,g}$}

\vskip 0.5cm
\centerline{\sl $^{a}$ SISSA, International School for Advanced Studies, via Bonomea 265, 34136 Trieste, Italy and}
\centerline{INFN Sezione di Trieste, via Valerio 2, 34127 Trieste, Italy}
\smallskip
\centerline{\sl $^{b}$ CENTRA, Departamento de Física, Instituto Superior Técnico – IST,}
\centerline{Universidade de Lisboa – UL, Avenida Rovisco Pais 1, 1049-001 Lisboa, Portugal}
\smallskip
\centerline{\sl $^{c}$  Departamento de F\'{\i}sica Te\'orica \& IPARCOS,} \centerline{Universidad Complutense de Madrid, 28040 Madrid, Spain}
\smallskip
\centerline{\sl $^{d}$ Instituto de F\'isica, Computaci\'on e Ciencias Aeroespaciais (IFCAE),} \centerline{Universidade de Vigo, E-32004 Ourense, Spain}
\centerline{\sl $^{e}$ Institute of Theoretical Physics, Jagiellonian University, \L{}ojasiewicza 11, 30-348 Krak\'ow, Poland}
\smallskip
\centerline{\sl $^{f}$Dipartimento di Fisica, Sapienza Università di Roma, Piazzale Aldo Moro 5, 00185, Roma, Italy}
\smallskip
\centerline{\sl $^{g}$ INFN, Sezione di Roma, Piazzale Aldo Moro 2, 00185, Roma, Italy}

\vskip 0.5cm
\centerline{\small\tt 
vboyanov@sissa.it, luisj.garay@ucm.es, alberto.garcia.martin-caro@uvigo.gal,}
\centerline{\small\tt gerardo.garciamoreno@uniroma1.it}

\vskip 1.cm
\centerline{\bf Abstract} \vskip 0.2cm \noindent

A recent work [Senovilla, Port. Math. 80 (2023)] has shown that accretion of matter by a Schwarzschild-de Sitter black hole can take this object beyond a critical limit where the event and cosmological horizons coalesce, transforming it into a collapsing universe resembling a Kantowski–Sachs cosmology. In these configurations, dubbed ``ultramassive spacetimes", future asymptotic infinity and the event horizon associated with the black hole cease to exist. In this work, we take the first steps toward exploring semiclassical gravity effects on such geometries. In particular, we examine the fate of Hawking radiation in these backgrounds and analyze the behavior of the renormalized stress-energy tensor within the Polyakov approximation. We also explore the effect of semiclassical backreaction on the evolution of the horizons, focusing on cases in the vicinity of the extremal limit, and on the subsequent universal collapse. 

\noindent

\end{titlepage}

\pagestyle{empty}
\normalsize 

\tableofcontents
\normalsize
\newpage
\pagestyle{plain}
\setcounter{page}{1}

\section{Introduction}
It has been shown~\cite{Hayward1993,Woolgar1999} that there is an upper limit on the area of spatially stable (in the sense of~\cite{Andersson2005,Andersson2007}) or outer stable (in the sense of~\cite{Hayward1993,Hayward1993b}) marginally future-trapped surfaces (MTS) for spacetimes satisfying Einstein equations with positive cosmological constant $\Lambda > 0 $ and for which the Dominant Energy Condition (DEC) holds. This stability requirement is tantamount to implying the presence of a black hole region  enclosed by the MTS ~\cite{Hayward1993,Hayward1993b}. The specific bound for those with spherical topology is:
\begin{align}
    A < \frac{4 \pi}{\Lambda}.
    \label{Eq:AreaBound}
\end{align}
As pointed out by Senovilla~\cite{Senovilla2022}, such a result is counterintuitive since naively one would expect that the area of the MTS should increase with the mass that comes from the external region, i.e., one can conceive situations in which there is a flux of mass through the MTS and, a priori, there is no obvious reason why the area cannot keep increasing without a bound. 

In fact, Senovilla showed in~\cite{Senovilla2022} that for some specific dynamical spherically symmetric configurations described by the Vaidya-de Sitter metric, a steady accretion of matter by the black hole can lead to a transition whereby the mass overcomes the effect of the cosmological constant, and the black hole transforms into a collapsing universe. In such a setup, no observer is able to avoid the future singularity, as the asymptotic region disappears. One can think of these spacetimes as if the black hole horizon was growing up until it reached the cosmological horizon, a moment at which the black hole swallows the whole external region and the full spacetime is converted into a collapsing cosmology with a singularity in the future of every observer. A similar analysis was also performed for an Oppenheimer-Snyder collapse with a cosmological constant. Additionally, in~\cite{Senovilla2023} it was shown that spacetimes containing spatially stable MTS with area approaching indefinitely the bound in Eq.~\eqref{Eq:AreaBound} exhibit some universal properties. The bound was refined in~\cite{Senovilla2026}, where the intriguing possibility of forming these ultramassive spacetimes without a cosmological constant is also discussed.

In this paper, we take the first step toward understanding quantum and semiclassical gravity effects in these spacetimes. Given the absence of asymptotic regions, we find it convenient to consider local observables, even for discussing the fate of Hawking radiation. In particular, we focus on the computation of the Renormalized Stress Energy Tensor (RSET), within the so-called Polyakov approximation, which captures only the $s$-wave modes while neglecting their backscattering. We characterize the conditions under which the geometry exhibits Hawking radiation, and analyze whether the shrinking of the horizon due to the Hawking evaporation process can prevent the formation of ultra-massive spacetimes. Finally, we consider the behavior of the RSET once the ultra-massive collapse has been formed and discuss the role of semiclassical effects in the collapsing cosmological phase. We find qualitative agreement with the conclusions of Ref.~\cite{Gao2023}, where it is shown that a sufficiently rapid particle production can cause a geometry which, in the absence of quantum effects, would simply undergo a bounce, to develop a future Big Crunch-like singularity.

The remain of the article is structured as follows. In Sec.~\ref{Sec:SphericallySymmetric} we present the families of geometries that we will be considering; in Subsec.~\ref{Subsec:KottlerDescription} we describe the spherically symmetric solutions to Einstein equations in vacuum with a positive cosmological and in Subsec.~\ref{Subsec:Vaidya-deSitter} we describe their formation through gravitational collapse. In Sec.~\ref{Sec:RSET} we describe the computation of the RSET; in Subsec.~\ref{Subsec:RSET_generalities} we present the approximations that we use to compute it in our setup and in Subsec.~\ref{Subsec:RSET_Calc} we describe how we fix its potential ambiguities including, in particular, the choice of state. In Sec.~\ref{Sec:HawkingRadition} we discuss the fate of Hawking radiation in ultramassive spacetimes. We review how the formation of a horizon with nonvanishing surface gravity generically leads to a flux of Hawking radiation in Subsec.~\ref{Subsec:Hawking_NutShell}, in particular we illustrate this for black holes formed in de Sitter, which are not asymptotically flat in Subsec.~\ref{Subsec:HawkingRad_Submassive}. We then discuss how the same kind of reasoning can lead to the Hawking phenomenon for ultramassive spacetimes that are accreting matter adiabatically in Subsec.~\ref{Subsec:HawkingRadTwoShells}. We give the first step towards self-consistent solutions in Sec.~\ref{Sec:PerturbativeBackreaction} by including perturbatively the backreaction and studying its effect for almost critical configurations in the near horizon region in Subsec.~\ref{Subsec:NearCriticality}, where the Hawking evaporation is specially relevant and the validity of our approximations is ensured, and on the evolution of the the fully trapped region in Subsec.~\ref{Subsec:TrappedRegion}. We conclude in Sec.~\ref{Sec:Conclusions} by summarizing the main results of the paper and discussing future lines of research and open questions. We have added two additional appendices that review some known results for convenience, as they can be useful for a reader that is not familiar with them or the approach we present here. In App.~\ref{App:deSitter} we discuss the quantization of real massless scalar field in de Sitter spacetime, and in App.~\ref{App:Flat_Collapse} we discuss the phenomenon of Hawking radiation in asymptotically flat Vaidya spacetimes, performing the explicit computation for a thin shell in the subsection~\ref{Subsec:ThinShell}.

%
 \paragraph*{\textbf{Notation and conventions.}}

We use the signature $(-,+,+,+)$  for the spacetime metric and we work in natural units in which $c= G_N = \hbar=1$. For the curvature tensors we use the conventions in~\cite{Wald1984},  i.e., $[\nabla_\mu, \nabla_\nu] V^\rho =: - R_{\mu \nu \sigma}{}^\rho V^\sigma$. 
\section{Spherically symmetric ultramassive spacetimes}
\label{Sec:SphericallySymmetric}
In this section, we begin by briefly reviewing the spherically symmetric vacuum solutions of Einstein equations with a positive cosmological constant, for which a straightforward generalization of Birkhoff's theorem constrains the form of the geometry and implies the existence of an additional Killing field~\cite{Eiesland1925,Stephani2003}, (see also~\cite{Schleich2009}). We then review the Vaidya-de Sitter metric, focusing in particular on the transition to ultra-massive spacetimes, following closely the discussion presented in~\cite{Senovilla2022}.

\subsection{Kottler (Schwarzschild de-Sitter) geometry}
\label{Subsec:KottlerDescription}
Consider the Kottler metric described by the following line element: 
\begin{align}
    \dd s^2 = - f(r) \dd t^2 + \frac{\dd r^2}{f(r)} + r^2 \dd \Omega^2,
    \label{Eq:KottlerMetric}
\end{align}
where $\dd \Omega^2 = \dd \theta^2 + \sin ^2 \theta \dd \varphi^2 $ represents the metric of the unit radius 2-sphere and the function $f(r)$ is given by:
\begin{align}
    f(r) = 1 - \frac{2m}{r}  - \frac{r^2}{\ell^2}.
    \label{Eq:f_Kottler}
\end{align}
It represents together with the Nariai solution~\cite{Stephani2003}\footnote{In fact, the Nariai solution can be obtained as a limit of spacetimes in the sense of~\cite{Geroch1969}, of the Kottler metric with critical values of the parameters. It corresponds to the case presented below in which $f(r)$ has only one positive double zero.}, which we will not discuss here, the only spherically symmetric solution to vacuum Einstein equations with a positive cosmological constant $\Lambda$, related to the parameter $\ell$ through $ \Lambda = 3 /\ell^2$. In fact, aside from spherical symmetry, it manifestly exhibits an additional Killing vector field $\partial_t$. In the case $m = 0$, it describes the static patch of   de Sitter geometry (whose Killing algebra is $\mathfrak{so}(1,4)$), whereas in the case $\Lambda = 0$ (which is equivalent to taking the $\ell \to \infty$ limit) we recover the standard Schwarzschild metric. For nonzero $m$ and finite $\ell$, we can introduce the critical mass $m_{\text{crit}} = \ell / (3 \sqrt{3})$. Depending on the ratio $m/m_{\text{crit}} $ (which in turn determines the number of real positive zeros of $f$) we have the following three scenarios: 
\begin{itemize}
    \item \textbf{Two zeros:} For $ m < m_{\text{crit}}$, $f(r)$ has two simple positive real roots at $r_{-} $ and $r_{+} $ (with $r_{-}(\ell, m) < r_{+}(\ell, m)$), whose analytic expressions can be found in App.~A of Ref.~\cite{Lake1977}. These represent the event and cosmological horizons, respectively. The function $f(r)$ is positive in the interval $ r \in (r_{-}, r_{+})$, this being the static patch analogous to the external region of a purely Schwarzschild black hole. In the maximal analytic extension of the metric, the interval $r\in(0,r_-)$ corresponds to a black hole trapped region in the future, and a white hole anti-trapped region in the past, with $r=0$ being a curvature singularity. The interval $r\in (r_{+},\infty)$, on the other hand, corresponds to the future and past cosmological regions, namely, the regions between the future cosmological horizon and $\mathscr{I}^{+}$ in the future, and between the past cosmological horizon and $\mathscr{I}^{-}$ in the past, respectively. The constant-$r$ spheres in the future (past) cosmological regions are anti-trapped (trapped). The maximal analytical extension has a repeating pattern with infinite repetitions of all these regions (although points can be identified  changing the topology of the maximal analytic extension of the  spacetime). A Penrose diagram of (part of) this maximal extension is depicted in Fig.~\ref{Fig:Kottler_submassive}. We note that for this case
    \begin{align}
        2m < r_{-} < 3m < r_{+} \qquad \text{and} \qquad r_{-} < \frac{\ell}{\sqrt{3}} < r_{+} < \ell.  
    \end{align}
\begin{figure}
    \centering
    \includegraphics[width=0.45\linewidth]{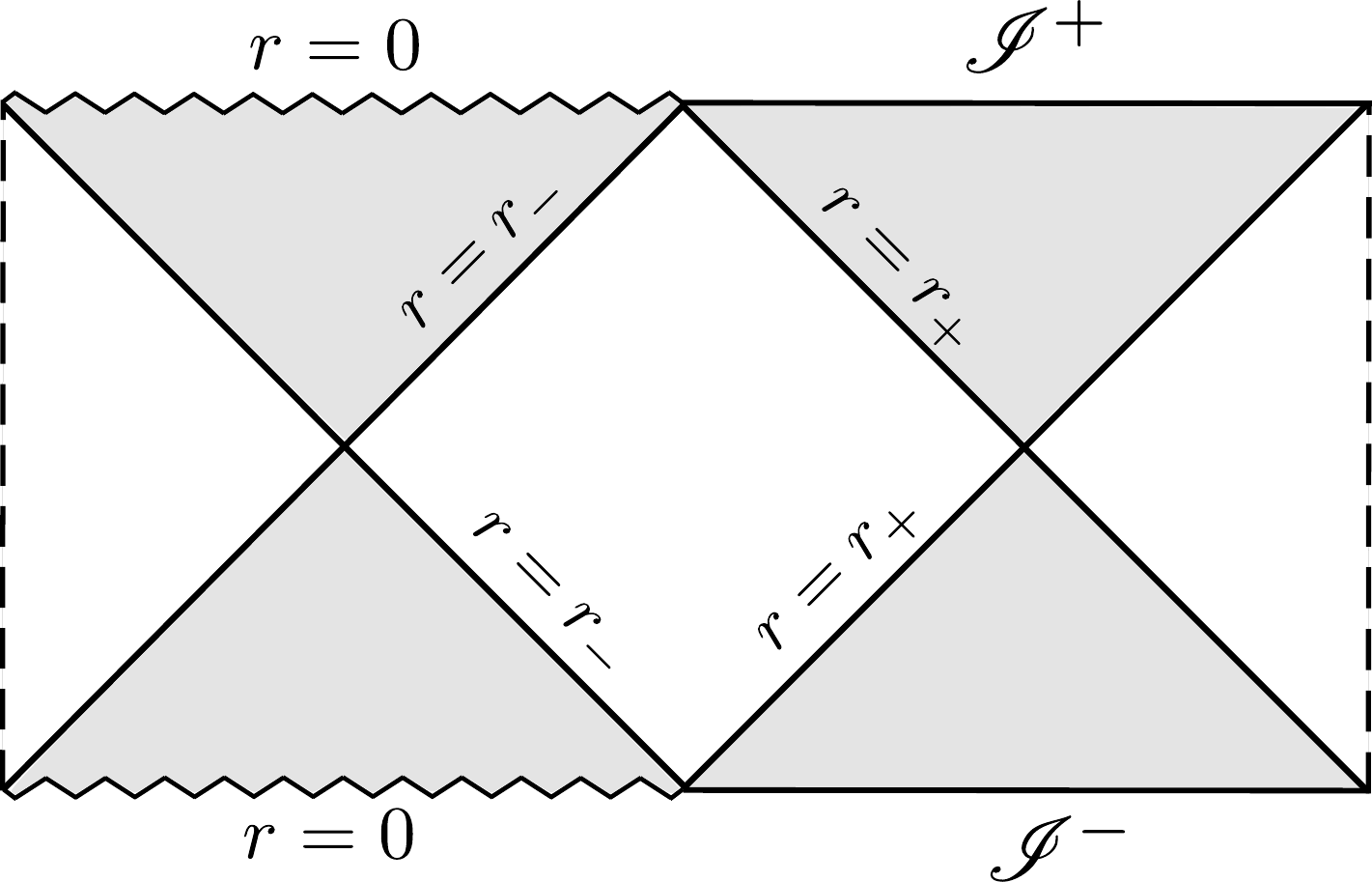}
    \caption{Penrose diagram of the maximal extension of a subcritical $m<m_\text{crit}$ Kottler spacetime. The shaded regions on the left represent the trapped and anti-trapped regions of the black hole (to the future) and white hole (to the past). The shaded regions on the right represent the past and future cosmological regions. The maximal extension of this geometry contains infinite copies of the represented patch, matched along the dashed lines, unless topological identifications are performed.}
    \label{Fig:Kottler_submassive}
\end{figure}
    \item \textbf{A single zero:} For $m=m_{\text{crit}} $, the function $f$ has a double positive root at \mbox{$r_0=3m=\frac{\ell}{\sqrt{3}}$}. We have the region $r\in(0,r_{0})$, with the singularity at $r = 0$, the degenerate horizon at $r=r_0$, and the exterior region $r\in(r_0,\infty)$. There are two possible spacetimes this construction can represent, one in which the asymptotic region is in the past and the singularity is in the future, and one in which the singularity is in the past and the asymptotic region in the future. We will consider the former case, as it is the one related to the dynamical formation of a black hole. The $r\to\infty$ limit therefore corresponds to past null infinity $\mathscr{I}^{-}$. The sphere at $r = r_0$ is marginally trapped, and all other $r=\text{const.}$ spheres are future trapped. The function $f$ is nonpositive and therefore, the vector $\partial_r$ is timelike everywhere except at $r = r_0$, where it becomes null. It is therefore more intuitive to relabel the coordinates as $t \to X$ and $r \to -T$ (with the minus sign making $T$ future-oriented), so that the metric reads
        \begin{align}
            \dd s^2 = - \frac{\dd T^2}{\abs{f(T)}} + \abs{f(T)} \dd X^2 + T^2 \dd \Omega_2^2, 
            \label{Eq:TX_coordinates}
        \end{align}
    where $T$ is a timelike coordinate, $X$ is spacelike (with $\partial_X$ a spacelike Killing field), and the function $f(T)$ is the one given in Eq.~\eqref{Eq:f_Kottler} with $r\to-T$. The maximal extension of the spacetime once again has a repeating pattern (again with possible identifications and matching). The Penrose diagram of part of this maximal extension is shown in Fig.~\ref{Fig:Kottler_critical}. 
\begin{figure}
    \centering
    \includegraphics[width=0.45\linewidth]{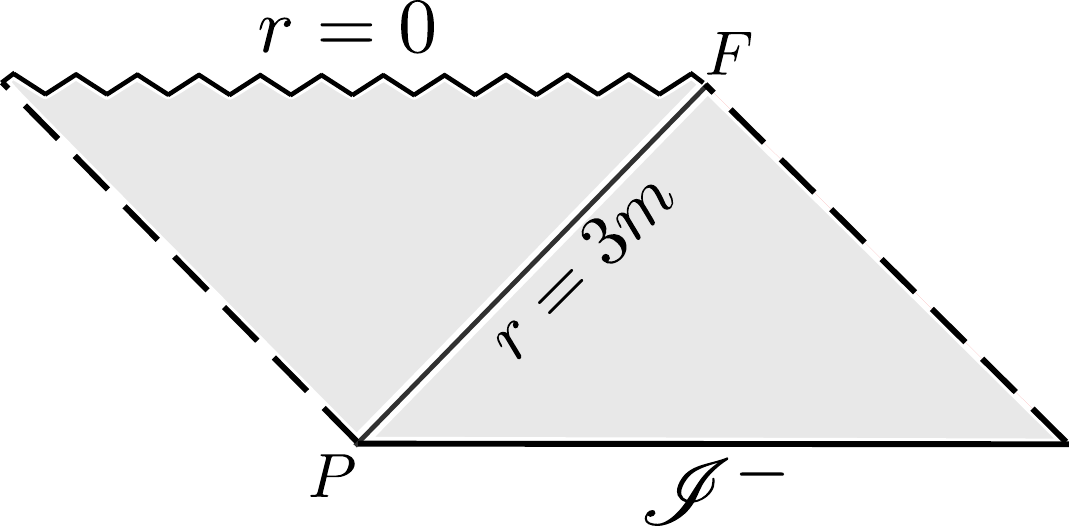}
    \caption{Penrose diagram for a Kottler spacetime in the critical case. All spheres are trapped, except at the interface between the past cosmological region and the future black hole region $r =3m$, where they are marginally trapped. The maximal extension again contains infinite copies matched along the dashed lines, unless topological identifications are performed.} 
    \label{Fig:Kottler_critical}
\end{figure}
    \item \textbf{No zeros:} For $m > m_{\text{crit}} $ the  function $f(r)$ does not have any positive real roots. This function is negative in the whole domain $r\in(0,\infty)$, making it once again convenient to relabel the coordinates to $(T,X)$, and to write the metric as Eq.~\eqref{Eq:TX_coordinates}. We also once again have the two possibilities of the spacetime either having a past asymptotic region and a future singularity, or a past singularity and a future asymptotic region. The former is again the one of physical relevance for black hole formation. In this case, we have $T\in(-\infty,0)$, with $-\infty$ corresponding to $\mathscr{I}^{-}$ and $0$ to a Big-Crunch-like singularity. This spacetime represents a collapsing universe whose spatial sections have topology $\mathbb{R} \times \mathbb{S}^2$, where all the constant-$X$ spheres are future trapped. To be more precise, we can think of the constant-$T$ slices as infinite cylinders with radius $|T|$ and symmetric along the $X$ direction, which are shrinking from infinite radius to vanishing radius as $T$ increases. The Penrose diagram is depicted in  Fig.~\ref{Fig:Kottler_supercritical}. In this case, the solution belongs to the Kantowski-Sachs families of contracting cosmologies~\cite{Kantowski1966,Stephani2003}. 
\begin{figure}
    \centering
    \includegraphics[width=0.45\linewidth]{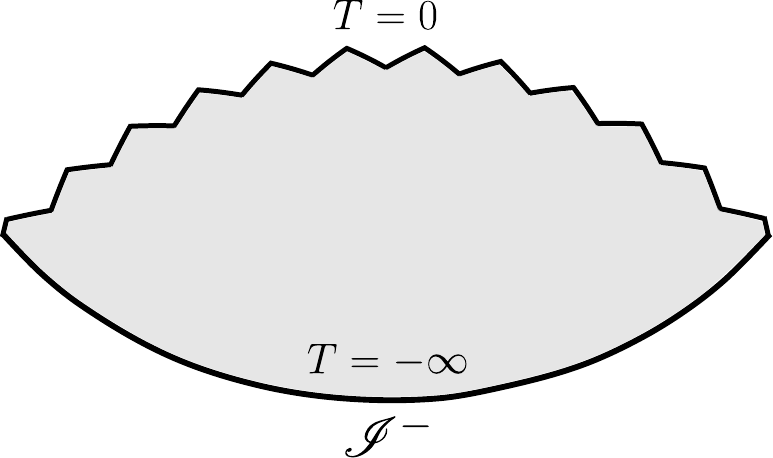}
    \caption{Penrose diagram for a Kottler spacetime in the super-critical, ultramassive case. All spheres are trapped and the geometry describes a Big Crunch cosmology.}
    \label{Fig:Kottler_supercritical}
\end{figure}
\end{itemize}

\subsection{Vaidya-de Sitter metric}
\label{Subsec:Vaidya-deSitter}
From a dynamical point of view, one can consider situations in which the accretion of matter allows the geometry to transition between the different configurations discussed above. In particular, we can conceive a case in which the black hole horizon grows dynamically due to matter accretion and eventually swallows the whole spacetime. This makes the asymptotic region disappear and the future region of the resulting geometry is akin to the third case (no zeros) described in the previous section. This means that all observers hit the future singularity in finite proper time.

This was illustrated to be a feasible physical scenario in Ref.~\cite{Senovilla2022} with both Vaidya-de Sitter and Oppenheimer-Snyder-de Sitter geometries. Here we focus on the Vaidya-de Sitter geometry, which is simpler for our considerations. The line element is given by 
\begin{align}
    \dd s^2 = - \left( 1 - \frac{2 m(v)}{r} - \frac{r^2}{\ell^2} \right) \dd  v^2 + 2 \dd v \dd r + r^2 \dd \Omega_2^2,
    \label{Eq:VaidyadeSitter}
\end{align}
where the range of the $(v,r)$ coordinates is $ - \infty < v < \infty$ and $ 0 < r < \infty$. The metric is a solution to Einstein equations with spherical ingoing null dust (pressureless) described by the stress-energy tensor
\begin{align}
    T_{\mu \nu} = \frac{2 m'(v)}{r^2} \delta_{\mu}^{\ v} \delta_{\nu}^{\ v},
\end{align}
where the prime represents differentiation of the single variable function with respect to its argument. For our purposes we will always take $m \geq 0$ and $m'\geq 0$, the second ensuring that the DEC is satisfied. We will consider an initially empty de Sitter spacetime that, due to the collapse of matter, transitions into one of the three situations described above, focusing in particular on the formation of ultramassive spacetimes, i.e., where the future asymptotic scenario is the third one (no zeros). We consider an $m(v)$ which grows from zero at $v=0$, to an asymptotic value $m_{\infty}$.

As a side note, since we are assuming $m(v) = 0$ for $v<0$, to avoid the formation of a naked singularity we need to impose that the mass function satisfies
\begin{align}
    \lim_{v \to 0^{+}} \frac{m(v)}{v} > \frac{1}{16},
    \label{Eq:BoundVaidya}
\end{align}
implying that the derivative of the function necessarily needs to be discontinuous~\cite{Kuroda1984,Wagh1999}. Physically this can be understood as requiring that the mass function profile grows fast enough so that   the growth of the apparent horizon is quick enough to hide the singularity inside of it. 

To understand the causal structure of this spacetime, we can look at the causal character of constant-$r$ hypersurfaces in different regions. The  normal one-form to these hypersurfaces is given by $ n_{\mu} = \delta^{r}_{\ \mu}$, and its norm is
\begin{align}
    n^2 =  \left(1 -  \frac{2m(v)}{r} - \frac{r^2}{\ell^2}\right).
\end{align}
As $r \to \infty$, $n^2 <0$, meaning that the constant-$r$ hypersurfaces are spacelike. Also, whenever $ m \neq 0$, $n^2 < 0$ in a neighborhood of $r = 0$. However, since we begin with zero mass in the past, there is always a region where these hypersurfaces are timelike. At any particular advanced time $v$, if $m(v) < m_{\text{crit}}$, these hypersurfaces will be timelike between $r_{-}(v)$ and $r_{+}(v)$, which depend on $v$ through their dependence on $m(v)$ which is now dynamical function. For $m_{\infty} \geq m_{\text{crit}}$, there exists a critical value $v_{\ast}$ such that $r_{-} (v_{\ast}) = r_{+} (v_{\ast})$. For $v > v_{\ast}$, $r$ becomes spacelike for all values of $v$.

The hypersurfaces defined by $r = r_{\pm}(v)$ represent a family of marginally trapped spheres. When $m'(v)\neq0$, the hypersurface $r = r_{-}(v)$ defines a spacelike marginally trapped tube (MTT) and $r = r_{+} (v)$ represents a timelike MTT. In Ref.~\cite{Senovilla2022} the former is dubbed the Dynamical Horizon (DH) and the latter the Timelike Membrane (TM), a nomenclature which we will respect in the present work. When $m_{\text{crit}} \geq m_{\infty}$, both merge for $v_{\ast}$ in a two-sphere that saturates the bound~\eqref{Eq:AreaBound}.

\begin{figure}
    \centering
    \includegraphics[width=0.6\linewidth]{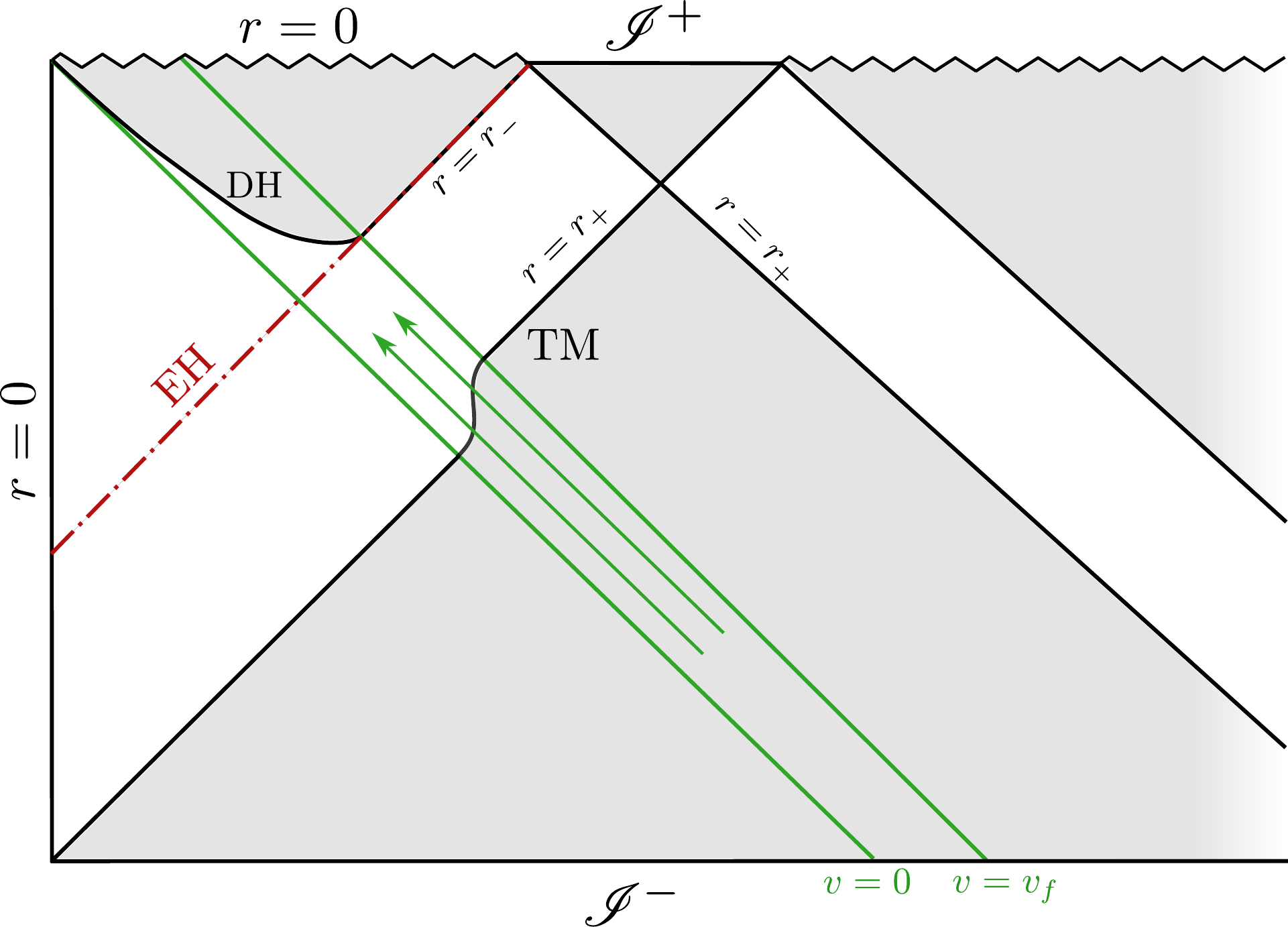}
    \caption{Penrose diagram for the formation of a black hole in de Sitter spacetime, with a sub-critical final mass $m_{\infty}< m_{\text{crit}}$. The shaded region on the top left represents the trapped region of the incipient black hole, while on the top right it represents a trapped region in the extension of the Kottler geometry. The middle two shaded regions are the past and future cosmological regions. The flux which collapses to a black hole is contained between $v=0$ and $v=v_f$.}
    \label{Fig:collapse_submassive}
\end{figure}

If $m_{\infty} < m_{\text{crit}}$, the spacetime represents the simple formation of a black hole in a de Sitter background, with the trapped and cosmological regions being disjoint. The Penrose diagram for this case is given in Fig.~\ref{Fig:collapse_submassive}.

\begin{figure}
    \centering
    \includegraphics[width=0.5\linewidth]{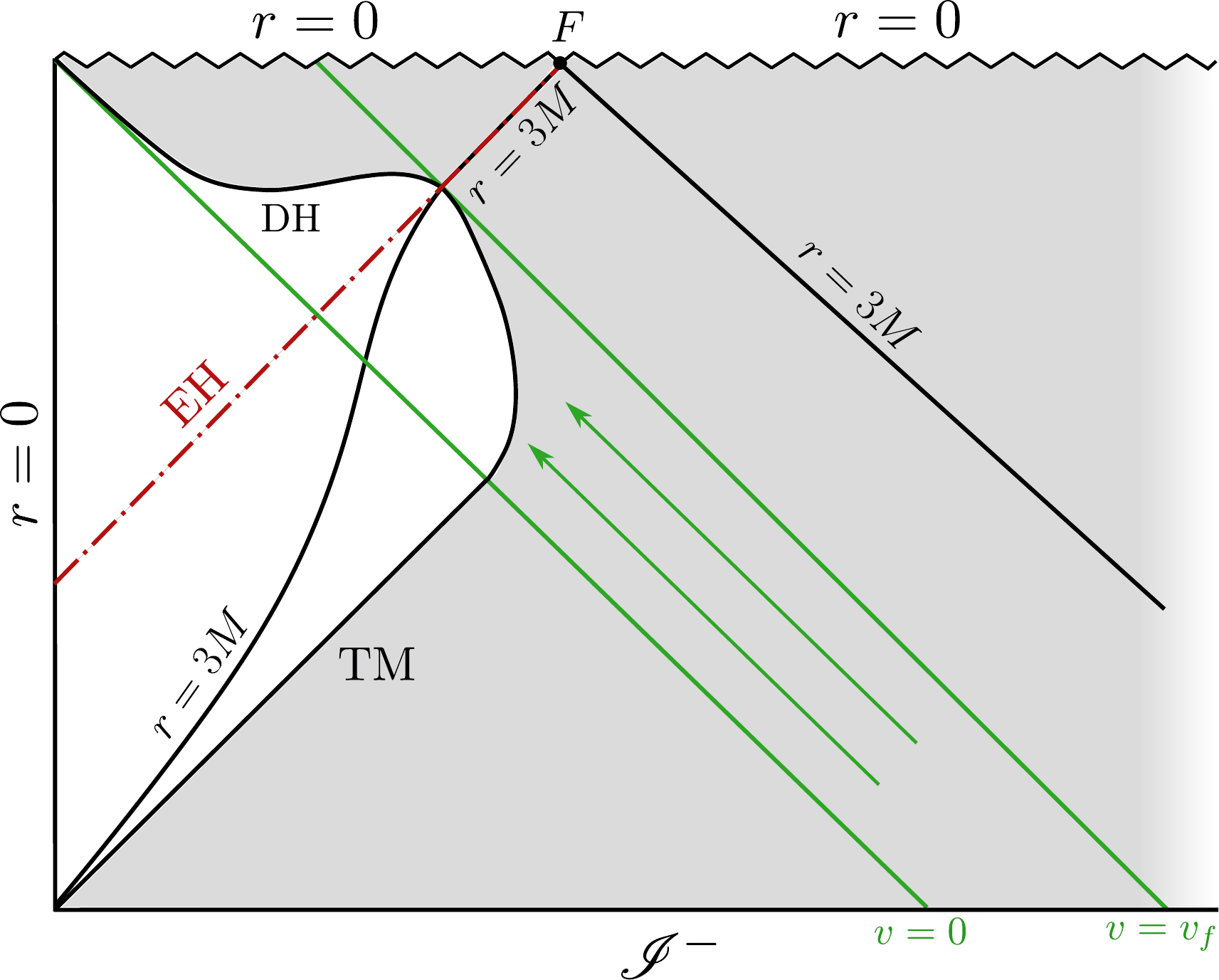}
    \caption{Penrose diagram for the formation of a black hole in de Sitter spacetime, with a critical mass $m_{\infty} = m_{\text{crit}}$. The shaded regions represent the black hole trapped region and the past cosmological region, which come into contact at the marginally trapped surface located at $r=3M$. Beyond $v=v_f$, the geometry is of the critical Kottler type, as in Fig.~\ref{Fig:Kottler_critical}. Note that the future asymptotic region already disappears in this case, and all the causal geodesics (except the null rays generating the horizon) hit the singularity in finite proper time.}
    \label{Fig:collapse_critical}
\end{figure}

On the other hand, if $m_{\infty} = m_{\text{crit}}$, we have the critical case for which the black hole apparent horizon and the cosmological horizon coalesce asymptotically. Fig.~\ref{Fig:collapse_critical} shows the Penrose diagram for this case, with a future region engulfed by a singularity except at a single point $F$ in the diagram which is a remnant of the future infinity region. The spacetime still exhibits a kind of event horizon whose null generator is future complete, as its orbit precisely ``ends'' at the point $F$ at infinity. Furthermore, very few special observers are able to reach this asymptotic region~\cite{Podolsky1999}. We notice that the corresponding figure in Ref.~\cite{Senovilla2022}, namely Fig. (4), represents a specific case in which the two surfaces: DH and TM only merge asymptotically at $F$.

\begin{figure}
    \centering
    \includegraphics[width=0.6\linewidth]{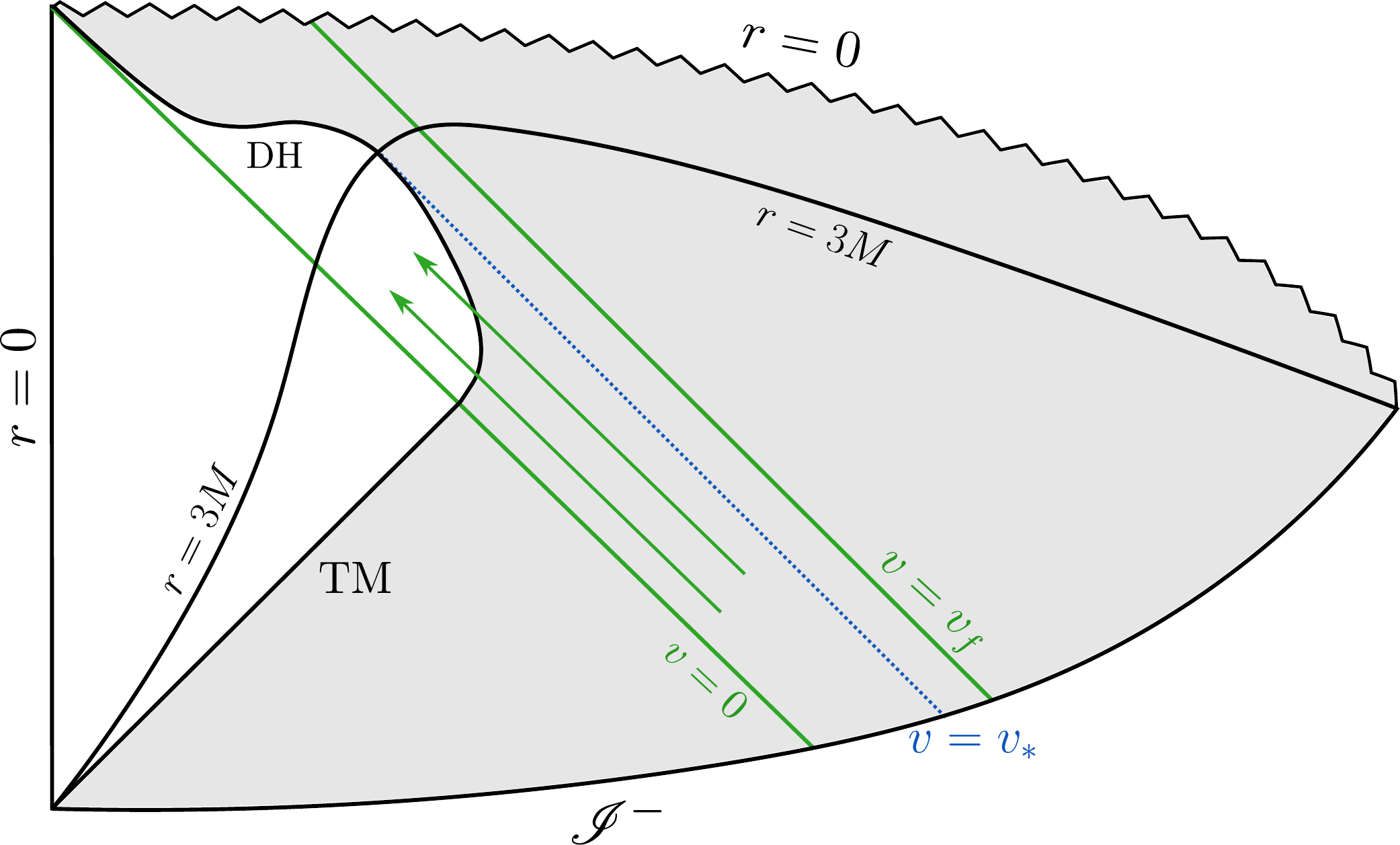}
    \caption{Penrose diagram for the formation of a black hole in the supercritical case $m_{\infty}>m_{\text{crit}}$. The shaded region covers the combination of the trapped and cosmological regions, ending in the future spacelike singularity. The cosmological and black hole dynamical horizons meet at $v=v_*$ and $r=3M$. In this case there are no causal curves which can avoid falling into the singularity. The resulting spacetime is inextendible and globally hyperbolic.}
    \label{Fig:collapse_ultramassive}
\end{figure}

Finally, the case $m_{\infty} > m_{\text{crit}}$ corresponds to the formation of an ultra-massive spacetime, the Penrose diagram of which is represented in Fig.~\ref{Fig:collapse_ultramassive}.

\begin{figure}
    \centering
    \includegraphics[width=0.6\linewidth]{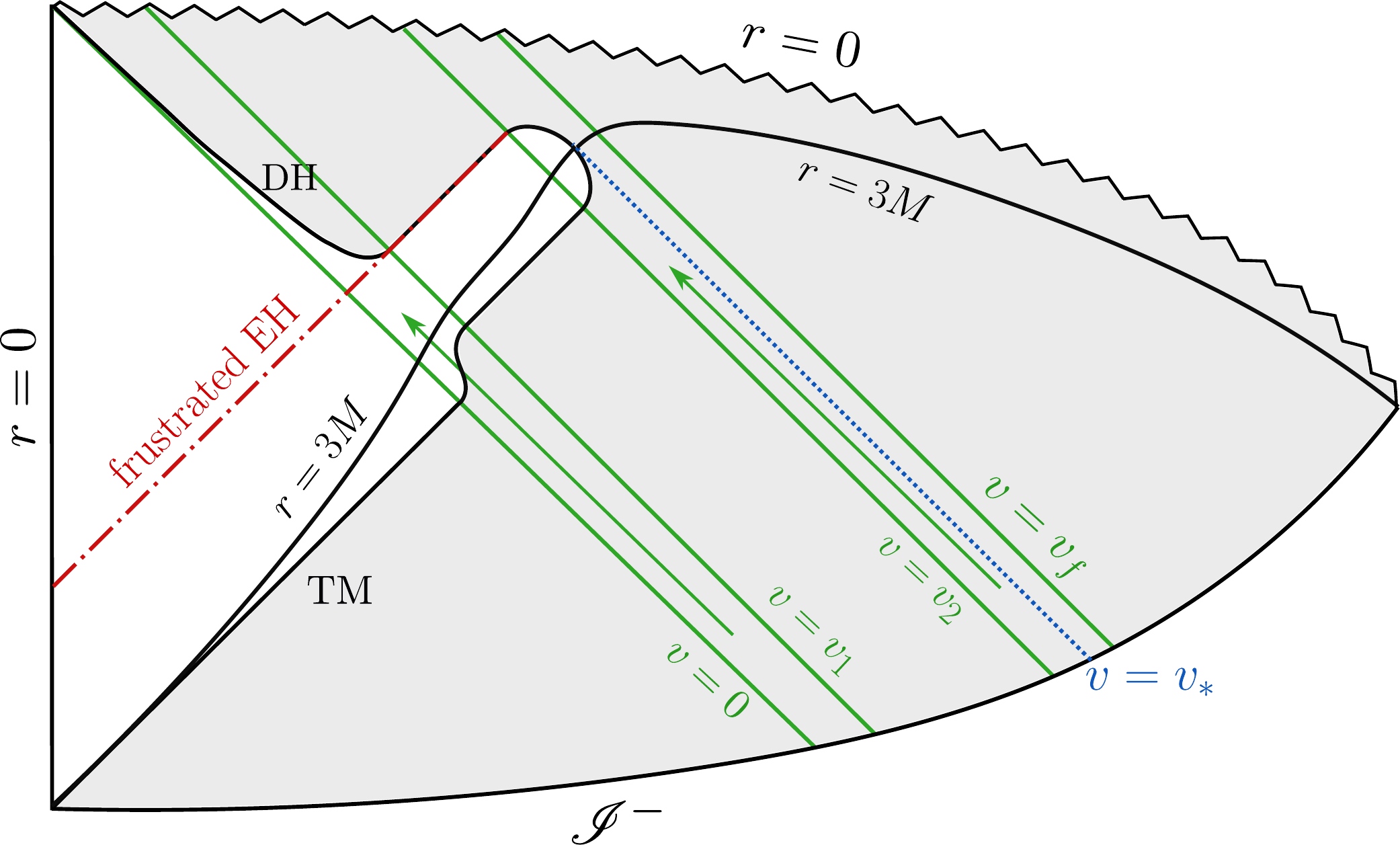}
    \caption{Penrose diagram for the formation of an ultramassive black hole from two pulses of infalling matter. Between $v=v_1$ and $v=v_2$, the configuration settles into a sub-massive Kottler geometry, and the dynamical horizon and timelike membrane become null surfaces. The critical mass is reached at $v=v_*$, during the accretion of the second pulse. We depict with a dot-dashed curve the null surface which would become an event horizon in the absence of the second pulse.}
    \label{Fig:TwoShells}
\end{figure}

As a final case, note that we can also consider a spacetime  for which $m'(v) \neq 0$ for two intervals of time $(0,v_{1})$ and $(v_2, v_f)$, with $0<v_1<v_2<v_f$, in such a way that $m(v_1) < m_{\text{crit}}$, but $m(v_f)>m_\text{crit.}$. Notice that this corresponds to the case in which a black hole is formed through the accretion of matter at $v = 0$, with an initial growth until $v_1$, followed by a static period until $v_2$, after which subsequent accretion occurs and the critical mass is reached at a time $v_{\ast} \in (v_2, v_f)$. The DH and the TM settle down, becoming null surfaces, until the second part of the infalling matter reaches it and it merges with the TM\footnote{This is dubbed a frustrated event horizon in~\cite{Senovilla2022}.}. The Penrose diagram is depicted in Fig.~\ref{Fig:TwoShells}.
\section{Asymptotically de Sitter spacetimes: RSET}
\label{Sec:RSET}

The goal of this work is to explore the behavior of quantum fields propagating on asymptotically de Sitter spacetimes in which a black hole forms dynamically and which subsequently become ultramassive, as described in the previous section ($m_\infty>m_\text{crit}$). For the sake of simplicity we will consider a free massless real scalar field propagating on top of these spacetimes. In particular we will target for the computation of the RSET. There are two main reasons for directly attempting to compute this object. First, given that in ultramassive spacetimes there is no future asymptotic region, it is not easy to follow a computation parallel to Hawking's and determine the spectrum of produced particles as measured by a distant observer. This difficulty does not arise for the RSET, as it is purely a local quantity, and therefore contains more information than the asymptotic computation of Bogoliubov coefficients. Second, computing the RSET is also advantageous as it is a first step toward understanding the backreaction of quantum vacuum energy on these spacetimes, which may become relevant and significantly alter the evolution of the system.

\subsection{Renormalized stress-energy tensor}
\label{Subsec:RSET_generalities}

Computing the RSET in four dimensional dynamical spacetimes is a task that is generally too difficult to achieve with current techniques. However, due to the spherical symmetry of the constructions we are exploring, we can resort to some approximations. Particularly, we use the Polyakov approximation which consists in two steps: i) taking the $s$-wave approximation, i.e., neglecting higher partial waves of the field, and, ii) neglecting the backscattering of the $s$-wave modes. The first approximation is well justified in spherically symmetric situations. For instance, most analysis agree that the biggest part of Hawking radiation is emitted in the $s$-wave sector~\cite{Fabbri2005}. The backscattering of the $s$-wave modes can, in general, be ignored only for specific geometries and some regions. For instance, for black holes in asymptotically flat spacetimes, it is only in the vicinity of the horizon, where the effective potential for the modes goes to zero, that the approximation is well justified.

Let us be mathematically more precise. Consider the following line element with the structure of a warped product with a spherically symmetric fiber:
\begin{align}
    \dd s ^2 = \tilde{g}_{ab} (x) \dd x^a \dd x^b + r^2 (x) \dd \Omega^2.
    \label{Eq:2D_metric}
\end{align}
The Klein-Gordon equation for a massless real scalar field propagating on top of it can be expressed as
\begin{align}
    \Box_{4D} \phi = \frac{1}{\sqrt{-\tilde{g}} r^2 } \partial_a \left( \sqrt{- \tilde{g}} r^2 \tilde{g}^{ab} \partial_{b} \phi \right) + \frac{1}{r^2} \Delta_{\mathbb{S}^2} \phi = 0,
\end{align}
where $\Delta_{\mathbb{S}^2} $ represents the Laplacian on the unit radius two-sphere $\mathbb{S}^2$. The first step of the Polyakov approximation amounts to neglecting any angular dependence on the field and hence eliminating the term with Laplacian on the unit two-sphere. With this, we can rearrange the equation as
\begin{align}
    \Box_{2D} \phi + \frac{2}{r} \tilde{g}^{ab} \left( \partial_a r \right) \left( \partial_b \phi \right)=0, 
    \label{Eq:ValidityPolakov}
\end{align}
where $\Box_{2D}$ is the D'Alambertian in the 2D metric. The second step of the Polyakov approximation consists of neglecting the last term of this equation. In the case of black hole setups, this term is inversely related to the value of the redshift of outgoing rays, so it indeed becomes negligible in the vicinity of horizons. 

With this approximation, the equation for the field is given by the 2D Klein-Gordon operator with respect to the metric $\tilde{g}_{ab}$ from Eq.~\eqref{Eq:2D_metric}. The massless 2D case can be solved explicitly taking advantage of the conformal symmetry. Given the line element expressed in double null coordinates,
\begin{align}\label{2d_null_metric}
    \dd s^2_{2D} = \tilde{g}_{ab} \dd x^a \dd x^b =  - C(u,v) \dd u \dd v, 
\end{align}
the expression for the two-dimensional RSET takes on the form~\cite{Fabbri2005}:
\begin{align}
    & T_{uu} = - \frac{\sqrt{C}}{12 \pi} \partial_u^2 \frac{1}{\sqrt{C}} + F(u), \label{Eq:Tuu}\\
    & T_{uv} = \frac{1}{96 \pi} \left( \frac{ 4 \partial_u C \partial_v C - 4 C \partial_u \partial_v C}{C^2} \right)  + \lambda C/2 , \label{Eq:Tuv} \\
    & T_{vv} = - \frac{\sqrt{C}}{12 \pi} \partial_v^2 \frac{1}{\sqrt{C}} + G(v),  \label{Eq:Tvv}
\end{align}
where all the ambiguity in the vacuum choice is encoded in the two functions $F(u)$ and $G(v)$, and $\lambda$ represents a finite constant associated with a local counterterm in the action that is tantamount to a renormalization of the cosmological constant in the 2D setting.

We note that the choice of vacuum state in this 2D setup can be related to a choice of a preferred pair of null coordinates $\{u,v\}$ for which $F=G=0$. Qualitatively this can be seen from the fact that a change of coordinates $u \rightarrow U(u)$ and $v \rightarrow V(v)$ has the same functional freedom as a choice of $F(u)$ and $G(v)$ for a fixed set of coordinates. 

More precisely, the quantization in two dimensions is performed by choosing a preferred basis of mode solutions to the 2D wave equation $\Box_{2D} \phi=0$. If this basis can be written as a set of plane waves in the coordinates $\{u,v\}$, then the RSET in the vacuum state of this quantization has the form \eqref{Eq:Tuu}-\eqref{Eq:Tvv} with $F=G=0$. If, on the other hand, the choice were a set of plane waves in a different pair of coordinates $\{U,V\}$, then the RSET in the vacuum state of this non-equivalent quantization, written in the coordinates $\{u,v\}$, would be given by the above expressions with the functions
\begin{align}
    F = - \frac{1}{12 \pi} \{ U, u \}_{\text{Sch}} \qquad G = - \frac{1}{12 \pi} \{ V, v \}_{\text{Sch}}, 
\end{align}
where the Schwarzian derivative is defined as
\begin{align}
    \{ y, x \}_{\text{Sch}}= \frac{y'''(x)}{y'(x)} - \frac{3}{2} \left( \frac{y''(x)}{y'(x)} \right)^2.
\end{align}
The 4D Polyakov RSET is then constructed from this 2D RSET by dividing the components of the latter by the area of spheres and identifying them with the components of the non-angular sector of the former, with the angular sector being set to zero:
\begin{align}
    T^{(4D)}_{\mu \nu} = \frac{1}{4 \pi r^2 (x) } T^{(2D)}_{ab} \delta^a_{\ \mu} \delta^{b}_{\ \nu}. 
    \label{Eq:4D_Tensor}
\end{align}
In practical terms, the choice of quantization reduces to the choice for the functions $F$ and $G$, and the constant $\lambda$, such that the corresponding vacuum state is physically well motivated. We note that fixing these functions and determining the RSET is extremely simple as compared to obtaining this quantity with other approximations, such as the $s$-wave approximation introduced in~\cite{Fabbri2005,Fabbri2005b}. The $s$-wave approximation requires specifying three functions of both $u$ and $v$, since the chirality of the effective conformal field theory is lost: these functions are constrained to satisfy two first order partial differential equations, making it much less straightforward to define the vacuum state.

In the next subsection we address the question of how to choose the functions $F$ and $G$. By considering simple, matter-free examples, we find a prescription for a natural vacuum state which we can subsequently use for any configuration in which the spacetime contains an asymptotically de Sitter part connected to $\mathscr{I}^{-}$.

\subsection{Vacuum choice in de Sitter and Kottler spacetimes}
\label{Subsec:RSET_Calc}

In scenarios of gravitational collapse, the past (null) infinity is usually regarded as a natural place for giving initial conditions which determine a field quantization. For massless fields in asymptotically flat spacetimes, a Minkowski-like quantization is typically performed at $\mathscr{I}^{-}$. Since the spacetimes we consider here are asymptotically de Sitter in the past, we will now turn our attention to determining a physically adequate quantization in asymptotically past de Sitter spacetime, which can then be applied to the situations of interest, in particular ultramassive spacetimes. In fact, this is in line with the role of $\mathscr{I}^{-}$ as a natural surface for prescribing initial data for Einstein equations with a positive cosmological constant~\cite{Friedrich1986}.

\begin{figure}
    \centering
    \includegraphics[width=0.4\linewidth]{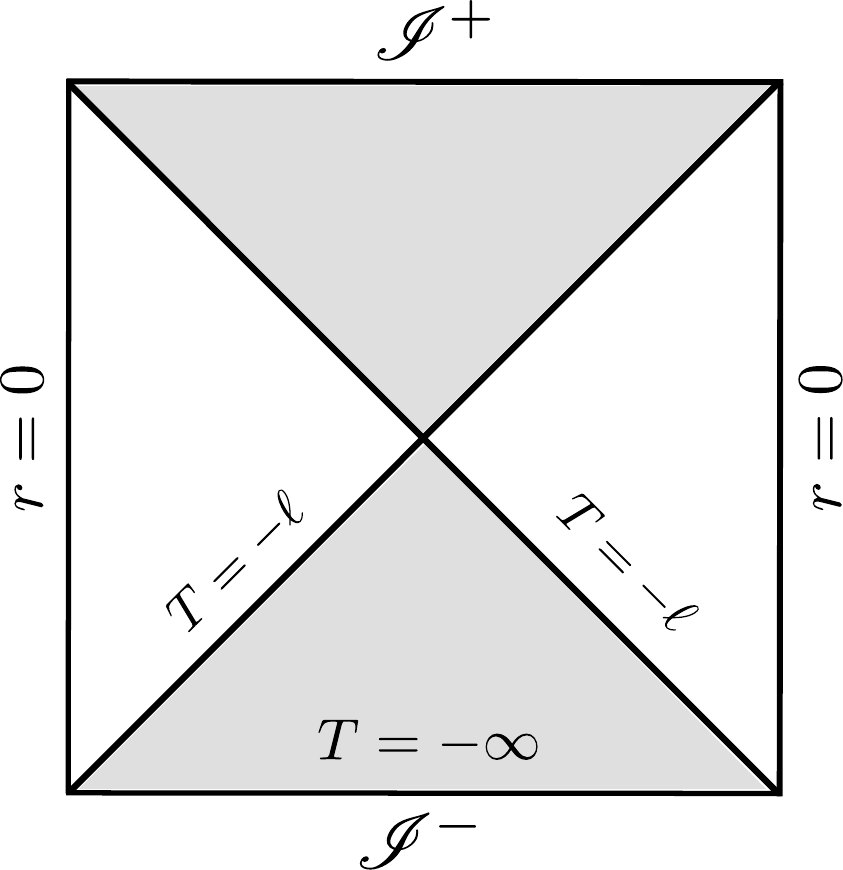}
    \caption{Penrose diagram of de Sitter spacetime. The shaded regions represent the past and future cosmological regions, where spheres are trapped and anti-trapped respectively. The solid lines at 45 degrees represent the cosmological horizons, and the vertical lines represent the south and north pole of the three spheres of the hyperboloid.}
    \label{Fig:deSitter}
\end{figure}

\noindent \paragraph*{\textbf{de Sitter spacetime.}} We start by considering the de Sitter metric, i.e., the line element in Eq.~\eqref{Eq:KottlerMetric} with the function $f(r)$ from Eq.~\eqref{Eq:f_Kottler} with $m = 0$. Furthermore, rather than the static patch, we will focus on the region beyond the cosmological horizon which is connected to $\mathscr{I}^{-}$, see Fig.~\ref{Fig:deSitter}. There the metric can be expressed as
\begin{align}
    \dd s^2 = - \frac{\dd T^2}{\abs{f(T)}} + \abs{f(T)} \dd X^2 + T^2 \dd \Omega_2^2.  
    \label{Eq:deSitterAsympt}
\end{align}
where the function $f(T)$ is given by Eq.~\eqref{Eq:f_Kottler} with $m =0$,
\begin{align}
    \abs{f(T)} = \frac{T^2}{\ell^2} - 1.
\end{align}
We recall that, for regions where $r$ is a timelike coordinate, we relabel our coordinates as $r\to-T$ and $t\to X$, as discussed in Sec.~\ref{Subsec:KottlerDescription}. From $\mathscr{I}^-$ to the past cosmological horizon, these coordinates span $T\in(-\infty,-\ell)$ and $X\in(-\infty,\infty)$.

As a first step toward constructing double null coordinates, we define the analogue $T_{\ast}$ of the tortoise coordinate for the now timelike $T$, which gives us
\begin{align}
        T_{\ast} = \frac{\ell}{2} \log \left( \frac{T- \ell}{T+\ell} \right),\qquad T = - \ell \coth \frac{T_{\ast}}{\ell},
        \label{Eq:taudesitter}
\end{align}
In this new coordinate $T_{\ast}$, past null infinity ($T\to-\infty$) corresponds to the finite $T_{\ast}\to 0$, while the past cosmological horizon ($T=-\ell$) corresponds to the limit $T_{\ast}\to\infty$. Using the inverse relation from \eqref{Eq:taudesitter} we can write the non-angular sector of the metric in $(T_{\ast},X)$ for this patch of the geometry as
\begin{align}
    \dd s_{2D}^2 = \frac{1}{\sinh^2 \left( \frac{T_{\ast}}{\ell} \right)} \left( - \dd T_{\ast}^2 + \dd X^2 \right).
    \label{Eq:taumetricdesitter}
\end{align}
We can easily go to double null coordinates from here by introducing $ u = T_{\ast} - X$ and $v = T_{\ast} + X$, which span the range $ u, v \in (-\infty, \infty)$ with the constraint $v+u >0$, where saturating this inequality corresponds to the limit of past null infinity. In these coordinates, the line element becomes
\begin{align}
    \dd s_{2D}^2 = - \frac{1}{\sinh^2 \left( \frac{u+v}{2 \ell} \right)}  \dd u \dd v .
\end{align}
Here $C(u,v) = - \sinh^{-2} \left( \frac{u+v}{2 \ell} \right) $, and a straightforward computation yields the 2D RSET
\begin{align}
    & T_{uu} = - \frac{1}{48 \pi \ell^2} + F(u), \label{Eq:Tuudesitter} \\
    & T_{uv} = - \frac{\sinh^{-2} \left( \frac{u+v}{2 \ell} \right) }{48 \pi \ell^2 } - \frac{\lambda}{2} \sinh^{-2} \left( \frac{u+v}{2 \ell} \right), \label{Eq:Tuvdesitter} \\
    & T_{vv} = - \frac{1}{48 \pi \ell^2} + G(v). \label{Eq:Tvvdesitter}
\end{align}
Now, we need to make a choice of vacuum state, i.e., we need to choose $F(u)$ and $G(v)$. In de Sitter spacetime, which is maximally symmetric, the expectation is that the full 4D RSET should be proportional to the metric itself, in order to preserve the symmetries of the spacetime. However, it is clear that, a priori, this is impossible within the Polyakov approach since the $s$-wave approximation breaks the symmetries of the spacetime. In fact, we note that we are singling out a family of 2D spheres with this foliation (see App.~\ref{App:deSitter} for a careful discussion). 

This will not be a problem for the purposes of this work, as the symmetries of the spacetimes that we consider next will reduce to only spherical symmetry and, as such, the foliation in spheres will be intrinsic. Furthermore, in the case of massless fields on de Sitter spacetime, it is well known that there is no vacuum state that is invariant under full group of isometries of the spacetime, since the renormalized two-point function breaks this invariance~\cite{Allen1985,Allen1987}. This seems to be related to a linear growth of the two-point function with time~\cite{Anderson2020,Anderson2022,Newsome2024}, which is also observed in the submassive Kottler geometry. However, it seems to be absent for the RSET, as it involves second derivatives of the two-point function. 

Therefore, while the example that we analyze here is not of importance for the 4D de Sitter spacetime itself, it is relevant for the spacetimes we will consider next. Our strategy will be to find a vacuum prescription  which makes the RSET identically zero in de Sitter spacetime. We expect that the same prescription applied to other spacetimes that only approach de Sitter asymptotically at $\mathscr{I}^{-}$, e.g., the Vaidya-de Sitter spacetime that we are considering which contains a nontrivial matter content, leads to a RSET that only vanishes asymptotically as $\mathscr{I}^{-}$ is approached. In this way, all the nonvanishing parts of the RSET arise purely due to the dynamical evolution of the geometrical system or, loosely speaking, its deviation with respect to the de Sitter geometry. The prescription we want for de Sitter spacetime is therefore $F(u) = G(v) = 1/(48 \pi \ell^2)$ and $\lambda = 1 /(24 \pi \ell^2)$, which results in $T_{ab} = 0$ in the entirety of this geometry.

\noindent \paragraph*{\textbf{A working example: Kottler metric.}} Let us now consider the horizonless Kottler metric, the line element of which has the form of Eq.~\eqref{Eq:deSitterAsympt} with the redshift function
\begin{align}
    |f(T)| = \frac{T^2}{\ell^2} - \frac{2 m}{T} -1.
\end{align}
We will furthermore choose the parameters such that there are no horizons ($ m > m_{\text{crit}}$). Even though the Polyakov approximation is not well-suited to this setup, since the potential is not negligible, it still serves as a proof of principle that the prescription introduced above yields a vanishing RSET at $\mathscr{I}^{-}$ which does not remain zero throughout the whole spacetime. In particular, it diverges as the singularity is approached, as expected. As described in Sec.~\ref{Subsec:KottlerDescription}, the timelike coordinate covers the entire geometry in the range $T\in(-\infty,0)$, between past null infinity at $T \to - \infty$, and the future curvature singularity at $T \to 0^{-}$. We can proceed in the same way as above and first introduce the tortoise-like $T_{\ast}$ coordinate (which reaches $\mathscr{I}^{-}$ at finite value),
\begin{align}
    T_{\ast} = \int_{-\infty}^{T} \frac{\dd T'}{f(T')}.
    \label{Eq:KottlerConfTime}
\end{align}
The 2D sector of the metric can be expressed as
\begin{align}
    \dd s^2_{2D} = f \left( - \dd T_{\ast}^2 + \dd X^2 \right),
\end{align}
where $f$ can be described in terms of $T_{\ast}$ through the inverse of \eqref{Eq:KottlerConfTime}. We can then once again go to double null coordinates $ u = T_{\ast} - X$ and $v = T_{\ast} + X$, giving us the metric
\begin{align}
    \dd s^2_{2D} = - f\, \dd u \dd v.
\end{align}
Using the chain rule we can compute the components of the 2D RSET in terms of the function $f$ and derivatives with respect to its argument $T$,
\begin{align}
    & T_{uu} = \frac{1}{96 \pi} \left[ f f'' - \frac{1}{2}(f')^2 \right] + F(u),\\
    & T_{uv} = - \frac{f''f}{96\pi } + \frac{\lambda}{2} f(T) ,\\
    & T_{vv} = \frac{1}{96 \pi} \left[ f f'' - \frac{1}{2}(f')^2 \right]  + G(v).
\end{align}
Writing it explicitly in terms of the coordinate $T$ we find
\begin{align}
    & T_{uu} = - \frac{m}{8 \pi  \ell^2 T} - \frac{1}{48 \pi  \ell^2}+\frac{m^2}{16 \pi  T^4} + \frac{m}{24 \pi  T^3} + F(u),\\
    & T_{uv} = -\frac{T^2}{48 \pi  \ell^4}+\frac{m}{12 \pi  \ell^2 T}+\frac{1}{48 \pi  \ell^2}-\frac{m^2}{12 \pi  T^4}-\frac{m}{24 \pi  T^3} + \frac{\lambda}{2} \left( \frac{T^2}{\ell^2} - \frac{2m}{T} -1 \right), \\
    & T_{vv} = - \frac{m}{8 \pi  \ell^2 T}-\frac{1}{48 \pi  \ell^2}+\frac{m^2}{16 \pi  T^4}+\frac{m}{24 \pi  T^3}  + G(v).
\end{align}
Following our above reasoning we need to once again fix $ F(u) = G(v) = 1/(48 \pi \ell^2)$ and $\lambda = 1/(24 \pi \ell^2)$, in order for the components to go to zero at $\mathscr{I}^-$, where $T\to-\infty$. However, we note that in this case the time dependence of the system makes the tensor nonvanishing as we move forward to finite values of $T$. In fact, the tensor becomes infinite as the singularity at $T \to 0^{-}$ is approached. To better interpret the behavior of the RSET, we can change to $(T,X)$ coordinates, for which:
\begin{align}
    & T_{TT} = - \frac{\ell^2 m T^3}{6 \pi  \left[ - \ell^2 T (2 m+T)+T^4\right]^2}-\frac{\ell^4 m^2}{24 \pi  \left[ - \ell^2 T (2 m+T)+T^4\right]^2}, \\
    & T_{TX} = 0, \\
    & T_{XX} = \frac{7 m^2}{24 \pi T^4} + \frac{m}{6 \pi T^3} - \frac{m}{3 \ell^2 \pi T}. 
\end{align}
The $TT$ component behaves as $ - \ell^2 m^2/(6 \pi T^5)$ as $T\to-\infty$, where it is vanishingly small but positive. On the other hand, it behaves as $-1/(96 \pi T^2)$ as $T \to 0$, which means that there is a negative energy density for sufficiently large $T$ (particularly, for $T>-(\ell^2m)^{1/3}$). The 4D tensor in this approximation can be obtained as a direct application of Eq.~\eqref{Eq:4D_Tensor}.

\noindent \paragraph*{\textbf{Choosing a vacuum in general.}} For a generic asymptotically de Sitter metric in the past (i.e. one in which there exists a coordinate set in which the metric approaches the de Sitter line element~\eqref{Eq:deSitterAsympt} as $T \to - \infty$) we have enough freedom to choose the functions and the renormalization of the cosmological constant in such a way that the RSET vanishes in this asymptotic region. We will choose such a vacuum from now on unless otherwise stated.

At this point, it is important to revisit the question of the validity of the Polyakov approximation, particularly in the situations of our interest, namely the collapse to ultramassive spacetimes. As mentioned above, the Polyakov approximation is based on neglecting the potential term in Eq.~\eqref{Eq:ValidityPolakov}. This term is only approximately zero near horizons and, in fact, diverges close to the past asymptotic region.

With this in mind, we note that the choice of a vacuum state for which the RSET vanishes in the past has two benefits: on the one hand, it ensures that the spacetime is also a self-consistent semiclassical solution in this region, and on the other hand, it ensures that the contributions from this region, where our approximation is not reliable, are suppressed. Our intention with this prescription for the vacuum choice is to isolate the terms which emerge in the RSET solely due to the dynamics of the ultramassive spacetimes, with a particular emphasis in the near-horizon behavior. Other choices of vacuum might be worth exploring in future works, together with improved approximations to the RSET.

\section{Hawking radiation in ultramassive spacetimes}
\label{Sec:HawkingRadition}
We now begin with the Vaidya-de Sitter line element, which we express as
\begin{align}
    \dd s^2 = - \left( 1 - \frac{2 m(v)}{r} - \frac{r^2}{\ell^2} \right) \dd  v^2 + 2 \dd v \dd r + r^2 \dd \Omega_2^2,
\end{align}
where we recall that the range of the $(v,r)$ coordinates is $ - \infty < v < \infty$ and $ 0 < r < \infty$. To set the stage for quantization and computation of the Polyakov RSET, we will express this line element in double null coordinates. For that purpose, we introduce a coordinate $u$ such that, with the function $r(u,v)$, we can write the 2D non-angular sector of the metric as
\begin{align}
     \dd s^2_{2D} = \left[ - \left( 1 - \frac{2 m(v)}{r} - \frac{r^2}{\ell^2} \right) + 2 \left( \frac{\partial r}{\partial v} \right) \right] \dd v^2 + 2  \left( \frac{\partial r}{ \partial u}\right) \dd u \dd v.
\end{align}
For $u$ to be a null coordinate we need to require that 
\begin{align}
    \frac{\partial r }{\partial v} = \frac{1}{2} \left( 1 - \frac{2m(v)}{r}  - \frac{r^2}{\ell^2} \right)
    \label{Eq:DrDv}
\end{align}
holds on $u=\text{const.}$ surfaces. This PDE fixes $r(u,v)$ up to an arbitrary function of $u$. Fixing this arbitrary function  amounts to the choice of a particular null coordinate. For the regions in which $m'(v) = 0$, we can fix $u$ so that it is symmetric with respect to $v$, i.e., so that it satisfies the equation
\begin{align}
    \frac{\partial r}{\partial u} = - \frac{1}{2} \left( 1 - \frac{2m}{r}  - \frac{r^2}{\ell^2} \right) \label{Eq:DrDu}
\end{align}
on $v=\text{const.}$ surfaces. In those regions the solution can be analytically obtained in terms of integrals of the redshift function. If we define the tortoise coordinate $r_{\ast}$ as in Eq.~\eqref{Eq:KottlerConfTime}
\begin{align}
    r_{\ast} = \int^{r} \frac{\dd r}{1 - \frac{2 m}{r} - \frac{r^2}{\ell^2}},
    \label{Eq:Tartaruga}
\end{align}
and then introduce the outgoing null coordinate $u$ defined in terms of $v$ and $r_{\ast}$ as
\begin{align}
    u = v - 2 r_{\ast},
    \label{Eq:utartaruga}
\end{align}
the metric can be written in double null coordinates as
\begin{align}
    \dd s^2 = - f(r) \dd v \dd u + r^2 (u,v) \dd \Omega_2^2,
\end{align}
where $r$, and hence $f(r)$, are implicitly given in terms of $(u,v)$ through Eqs.~\eqref{Eq:Tartaruga}-\eqref{Eq:utartaruga}. The integral in Eq.~\eqref{Eq:Tartaruga} can be performed analytically, and the result can be expressed as a simple sum of logarithms using the solutions for the zeros of $f(r)$, although this will not be needed for our purposes. 

The $u$ coordinate obtained in this way for the region below (above) the matter content, i.e., for the region $v \in (-\infty, 0)$ (respectively, $v \in(v_f,\infty)$) is called $u_{<}$ (respectively, $u_{>}$) and is determined up to an integration constant. Once the constant is fixed in the ``in'' region, the function $r(u_{<}, 0)$ can be used as initial data to integrate the equation across the region where $m'(v) \neq 0$. This determines $r(u_{<}, v_f)$ and thereby provides a matching between the $u_{<}$ and $u_{>}$ coordinates. Once the metric has been completely expressed in double null coordinates $(u_{<},v)$, the computation proceeds straightforwardly along the lines of our discussion in Sec.~\ref{Sec:RSET}.

Although the full computation can be carried out numerically, for our purposes it is more instructive to employ an approximation that yields analytic results and makes the underlying physics more transparent. In Subsec.~\ref{Subsec:Hawking_NutShell} we show, within an adiabatic approximation, that any metric whose redshift function vanishes, and therefore induces an exponential peeling of nearby light rays, gives rise to an approximately thermal Hawking flux for the vacuum states under consideration. We follow the discussion of Ref.~\cite{Barcelo2010}. Since this is a well-established result, our discussion is intended primarily as a brief review and to establish the notation used throughout the rest of the paper. For comparison, in App.~\ref{App:Flat_Collapse} we present the analytic derivation of Hawking radiation in the more familiar toy model of a thin shell collapsing in an asymptotically flat Vaidya spacetime, illustrating how the same approximation successfully captures the essential features of the Hawking effect, as an analytic solution is available there.

We then turn our attention to the two scenarios of interest for us addressed in Subsecs.~\ref{Subsec:HawkingRad_Submassive}--\ref{Subsec:HawkingRadTwoShells}: the case in which a black hole ($m < m_{\text{crit}}$) is formed, and the case in which an ultramassive spacetime is created ($m>m_{\text{crit}}$). 

\subsection{Exponential peeling of null geodesics.}
\label{Subsec:Hawking_NutShell}
We will now show the derivation of the Hawking effect under the adiabatic approximation. For any function $m(v)$ that satisfies $m'(v)\neq 0$ only in a finite interval $(0,v_f)$, with $m(0)=0$ and $m(v_f)>0$, an approximate analytic treatment can be performed. This is due to the fact that the leading-order contribution originates from the peeling of null geodesics near the black hole outer apparent horizon and is therefore largely independent of the details of the mass function. The dominant semiclassical effect, as pertains to the evolution of the trapped region, stems from the logarithmic relation between the ``in" null coordinate $u_{<}$ and the ``out" one $u_{>}$. This is nicely illustrated by the toy model discussed in Appendix~\ref{App:Flat_Collapse}, where an analytic treatment is possible and allows for a thorough comparison.

To capture this logarithmic behavior, we perform an expansion around the outgoing null geodesics which remain in the vicinity of the horizon for a long time interval, as measured by the advanced time $v$. To this end, it is more convenient to work with the more general line element
\begin{align}
    \dd s^2 = - f(v,r) \dd v^2 + 2 \dd v \dd r + r^2 \dd \Omega^2.
\end{align}
We assume that for some values of $v$ a dynamical black hole outer apparent horizon is present and located at $r = R_h (v)$, defined by the conditions $f(v,r)|_{R_h} = 0$ and $\partial_rf(v,r)|_{R_h}>0$. We expand the redshift function close to the horizon to leading order:
\begin{align}
    f(v,r) =  \partial_r f(v,r) \Big\rvert_{r = R_{h}(v)} \delta r (v),
\end{align}
where $\delta r (v)=r-R_h(v)$. We can introduce the local surface gravity
\begin{align}
    \kappa(v) := \frac{1}{2} \partial_r f(v,r) \Big\rvert_{r = R_{h}(v)},
\end{align}
and using the series expansion of $f$ in the equation for outgoing null geodesics \mbox{$\dd r/ \dd v=f/2$}, we can write this equation as
\begin{align}
    \partial_v \delta r = \kappa(v) \delta r  - R_h'(v) + \order{\delta r^2}.
\end{align}
The solution for $\delta r(v)$ on these $u=\text{const.}$ geodesics up to leading order is given by
\begin{align}
    \delta r (v) = \delta r_0 e^{\int^v \dd v' \kappa(v')} -R_h(v).
\end{align}
For light rays which find themselves in the vicinity of the horizon (such that this expansion is valid) at late times, the dominant part in the exponent comes from the time interval for which $\kappa = \kappa_f=\text{const.}$, this being the surface gravity of the black hole once the configuration has settled. Therefore, for these light rays we have
\begin{align}
    \delta r \approx \delta r_0 e^{\kappa_f v}.
    \label{Eq:Exp_Peeling}
\end{align}
After this exponential peeling, as the light rays escape the gravitational well of the black hole, their radial separation transitions to being approximately constant in $v$ (while they are still far from any cosmological horizon). The same is true for the region before the matter collapse. In these regions, the $u_{<}$ and $u_{>}$ coordinates on a $v=\text{const.}$ slice are approximately proportional to the radius $r$. Given the dominant part of the radial dispersion~\eqref{Eq:Exp_Peeling}, these $u$ coordinates are related through
\begin{align}
    u_{>} \sim - \frac{1}{\kappa_f} \log \left( u_0 - u_{<} \right),
    \label{Eq:u_inout}
\end{align}
where $u_0$ is the horizon generator, c.f. Eq.~\eqref{Eq:ThinShellinout} where $u_{<}\to -4m_{\infty}$ as $u_{>} \to \infty$. From this, we can find the leading order of the outgoing flux at infinity, which expressed in the natural ``out'' coordinates there becomes
\begin{align}
    T_{u_{>}u_{>}} = \frac{\kappa_f^2}{48 \pi}. 
\end{align}
Furthermore, from this expansion, we can also find the conformal factor at leading order in $\delta r$ around the horizon to be $C(u,v) \approx e^{\kappa_f v}$, which leads to an ingoing flux of negative energy at the horizon, namely
\begin{align}
    T_{vv} = - \frac{\kappa_f^2}{48 \pi}. 
\end{align}
Notice that the main approximations involved here are that, once the object acquires a mass of order $1/\kappa_f$, the change in the surface gravity should be adiabatic with respect to this scale, $\partial_v {\kappa}/ \kappa_f^2 \ll 1$, as discussed in detail in~\cite{Barcelo2010}. In general, we can neglect any deviations from the Hawking flux arising from the details of a standard collapse, as these are highly suppressed~\cite{Mathur2009}. Furthermore, once the configuration has approximately reached its final mass, the adiabatic condition ensures that the subsequent accretion of  matter does not change the described picture.

\begin{figure}
    \centering
    \includegraphics[width=0.4\linewidth]{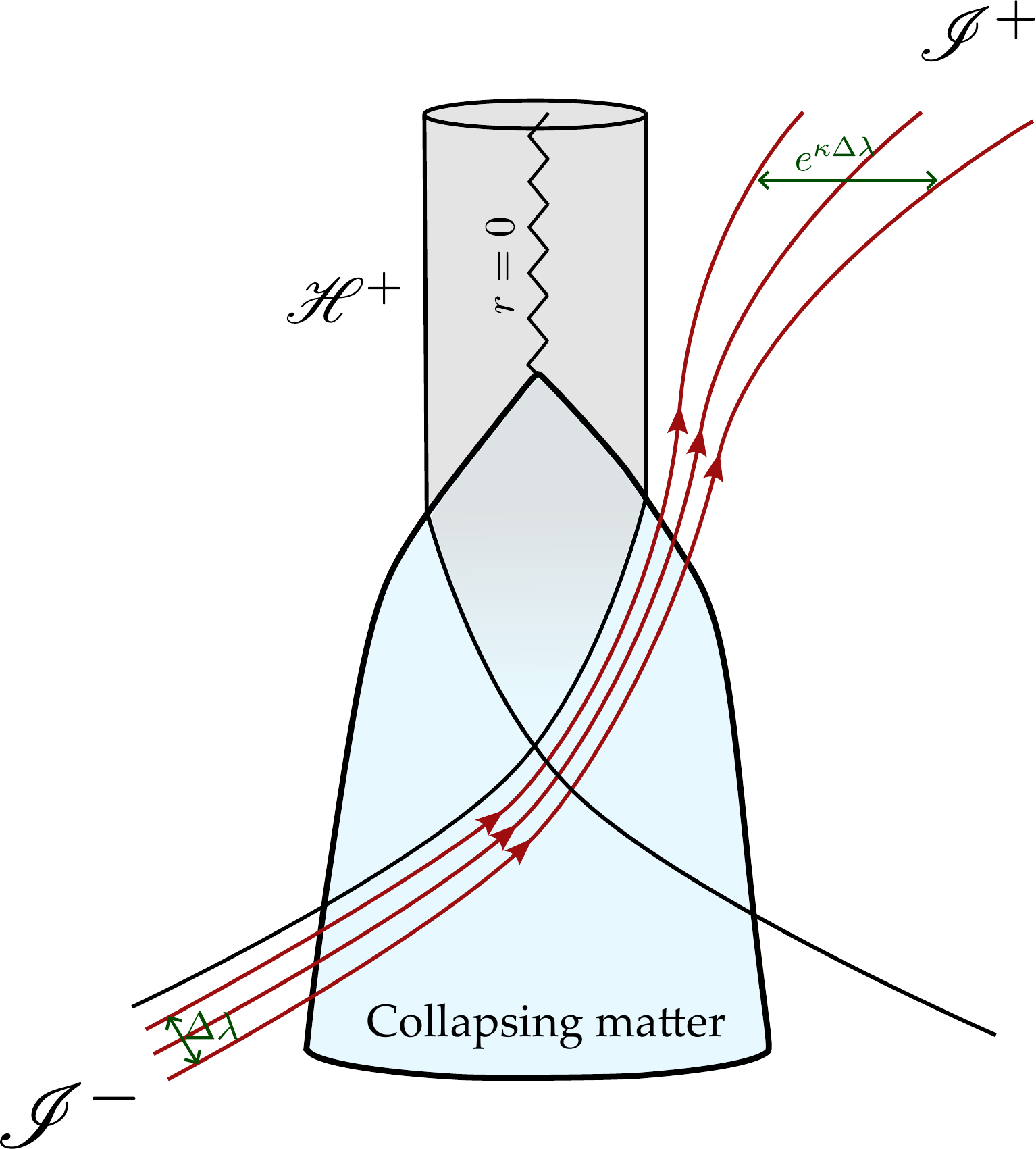}
    \caption{Pictorial representation of the matter infalling a black hole. The key ingredient for the Hawking radiation is the exponential peeling of the light rays, i.e., ingoing rays that are initially separated an affine distance $\Delta \lambda$, are exponentially separated at the end due to the gravitational time delay that they suffer when escaping. Additionally, they will be exponentially redshifted.}
    \label{Fig:PictorialHawkingFlux}
\end{figure}

In mathematical terms, this means that the requirement for the existence of a Hawking flux is that the distance (in affine parameter) between outgoing light rays is exponentially increased due to their scattering by the incipient black hole, see Fig.~\ref{Fig:PictorialHawkingFlux} for a pictorial representation.

\subsection{Submassive case}
\label{Subsec:HawkingRad_Submassive}

For the sake of simplicity, we consider a black hole formed by the collapse of a distributional matter profile consisting of an ingoing null thin shell, meaning that $m(v) = m \Theta (v)$ and $m < m_{\text{crit}}$. In other words, for $v<0$ the geometry is a patch of de Sitter spacetime, and for $v >0$ it corresponds to a patch of Kottler spacetime with $m < m_{\text{crit}}$. Its causal diagram simply corresponds to the one in Fig.~\ref{Fig:collapse_submassive}, but the region with nontrivial matter content $v \in (0,v_f)$ collapses to the single null ray $v = 0$ where the shell is supported.

The fact that there is an event horizon with nonzero surface gravity in the Kottler region already means that the geometry exhibits an exponential peeling of light rays. Given the above discussion, we can conclude that an outgoing flux of radiation arising from this exponential peeling is to be expected. Mathematically, this is precisely captured by the fact that there is a logarithmic divergence in the relation between the $u$ coordinates above and below the shell, as in Eq.~\eqref{Eq:u_inout}. 

To be more precise, consider the tortoise coordinate introduced in Eq.~\eqref{Eq:Tartaruga}. In our case, it will be piecewise defined as for $v < 0 $, where $f(r) = 1 - r^2 /\ell^2 $, we have $r^{\ast}_{<} (r)$ and for $v > 0$ we have $r^{\ast}_{>} (r)$ where $f(r) = 1 - 2m/r - r^2 /\ell^2$. This leads to two null coordinates $u$ defined as $u_{<} = v - 2 r^{\ast}_{<}$ and $u_{>} = v - 2 r^{\ast}_{>}$, valid for $v<0$ and $v>0$, respectively. In terms of them, we have that the metric is given by:
\begin{align}
    \dd s^2 = \begin{cases}
    - f \left( r^{\ast}_{<} (r) \right) \dd v \dd u_{<}  + r^2 \dd \Omega_2^2, \qquad \text{for} \qquad v<0, \\
    - f \left( r^{\ast}_{>} (r) \right)  \dd v \dd u_{>} + r^2 \dd \Omega_2^2, \qquad \text{for} \qquad v>0. 
    \end{cases}
\end{align}
To carry on with the quantization we need to obtain $u_{>}$ as a function of $u_{<}$. The matching between the two coordinates can be done following the steps sketched in the previous section. We need to expand the metric near the horizon and match both coordinates, which leads to the logarithmic relation in Eq.~\eqref{Eq:u_inout}, where $u_0$ represents the generator of the horizon. 

Extending this coordinate into the black hole region, the metric is given by
\begin{align}
    \dd s^2 = -C(u_{<},v) \dd u_{<} \dd v + r^2 \dd \Omega^2,
\end{align}
with $C(u_{<},v)$ having the piecewise expression
\begin{align}
    C(u_{<},v)  \begin{cases} 
      = 1 - \frac{r^2(u_{<},v)}{\ell^2}, & v < 0, \\
      \approx \left(1 - \frac{2m_{\infty}}{r(u_{<},v)} - \frac{r^2 (u_{<},v)}{\ell^2} \right) \frac{1}{\kappa_{-} (u_0-u_{<})},  & v > 0, 
   \end{cases}
\end{align}
where $\kappa_{-}$ is the surface gravity of the black hole horizon, $r$ is implicitly defined in terms of $(u_{<},v)$ through the tortoise coordinate, and the approximation in the second line indicates the leading order of an expansion around the horizon in $u_{<}-u_0$ (the region of validity of which expands radially along with outgoing geodesics). From this conformal factor, we can compute the RSET from Eqs.~\eqref{Eq:Tuu}--\eqref{Eq:Tvv}, with the derivatives of $r$ computed with the chain rule.

Notice that the RSET receives a constant contribution in both regions from the derivatives of the $1-r^2/\ell^2$ terms in the conformal factor that cancel with functions $F$ and $G$ associated with our vacuum choice. In addition, it receives a contribution from the $1/ \kappa_{-}(u_0-u_{<})$ term which precisely corresponds to the Hawking flux. Indeed, the $uu$ component reads
\begin{align}
    T_{u_{<} u_{<}} = \frac{1}{4 8 \pi (u_0 - u_{<} )^2 } \left[ 1 + \order{u_0-u_{<}} \right].
\end{align}
Note that here we are taking the late time behavior, i.e., its behavior on null surfaces $u_{<} = \text{const.}$ as they approach the horizon. We can change to $u_{>}$ coordinates and obtain the same flux as the one we obtain at $\mathscr{I}^{+}$ in the asymptotically flat case:
\begin{align}\label{tuu_h}
    T_{u_>u_>} \approx \frac{\kappa_{-}^2}{48 \pi}.
\end{align}
It is worth pointing out that in the Kottler case, there is no significant redshift or blueshift that this radiation suffers at the moment of crossing the future cosmological horizon, as can be surmised by the fact both the $u_<$ and $u_>$ coordinates are regular at this surface. The outward flux of Hawking radiation as such is thus a physically robust result well into the cosmological region.

\subsection{Ultramassive case}
\label{Subsec:HawkingRadTwoShells}

Consider a configuration in which the mass of the collapsing matter ultimately surpasses the critical limit, \mbox{$m_{\infty} > m_{\text{crit}}$}. As pertains to Hawking radiation, such a configuration can be reached in two qualitatively distinct ways. The first is the case in which \mbox{$m'(v) \gtrsim m_{\text{crit}}^2$} throughout the evolution, from the formation of a trapped region to surpassing the ultramassive limit. In such a case, the dynamical nature of the spacetime $m'(v) \neq 0$ will naturally lead to particle production, the details of which will depend on the particular black hole formation and accretion dynamics. However, due to the lack of a stationary or quasi-stationary period, the arguments used above for the derivation of a thermal Hawking-like flux would no longer be applicable. Consequently, the spectrum of the radiation produced would generally lack any resemblance to thermality, and its intensity would be dictated by the magnitude of $m'(v)$ and higher derivatives.

The second scenario is one in which a black hole is initially formed with some mass $m_i \lesssim m_{\text{crit}}$, after which a period of slow accretion follows, such that $m'(v) \ll m_{i}^2$. This adiabatic regime continues long enough for the black hole to thermalize, i.e. a period greater than the light crossing time, such that by the above arguments there appears a Hawking flux. We emphasize that this adiabatic regime is necessary for the emission of Hawking radiation. After this period, an ultramassive configuration is ultimately reached, either through the slow progressive accretion (which would gradually decrease the temperature and intensity of the Hawking radiation, as the surface gravity of the horizon decreases) or through a second non-adiabatic matter collapse. For simplicity, we can work with the latter case, and model the double collapse with just a pair of ingoing null shells. Starting with a de Sitter geometry for $v<v_1$, the first shell has a mass $m_1 \lesssim m_{\text{crit}}$. The corresponding sub-critical Kottler geometry lasts for a period $v_1<v<v_2$, during which thermal Hawking radiation is emitted. Finally, at $v=v_2$ the second shell falls in, and its mass $m_2$ is such that $m_{\infty} = m_1 + m_2>m_{\text{crit}}$, so that at later times the spacetime is an ultramassive configuration. The causal diagram in this case corresponds to the one in Fig.~\ref{Fig:TwoShells}, where the region that is not in vacuum will collapse to two single null lightrays where the shells will be supported.

As mentioned, the time separation between the shells $v_2-v_1$ needs to be greater than the characteristic time scale associated with the Hawking flux, this being of the order of the light crossing time of the black hole $\sim \kappa_-^{-1}$, where $\kappa_-$ represents the surface gravity of the black hole horizon associated with the first configuration. Under this assumption, following exactly the steps taken in the previous subsection, we find that the generated flux is
\begin{align}
    T_{u_>u_>} = \frac{\kappa_{-}^2}{48 \pi}.
\end{align}
Analogously to how this flux suffers no substantial blueshift after crossing the cosmological horizon in the sub-massive case, this is again true when it crosses the second shell at $v=v_2$ and enters the ultramassive region. The main difference is that in this case, instead of escaping to the asymptotic region, which in this case is absent, the flux enters the ultramassive region and later falls into the future singularity.

\section{Perturbative backreaction}
\label{Sec:PerturbativeBackreaction}

In this section we provide a first analysis of how semiclassical effects may modify the evolution of the geometry of an incipient ultramassive spacetime. In particular, we focus on two different regimes. Firstly, we will look at the dynamics of the black hole and cosmological horizons when an ultramassive configuration is approached in a sufficiently adiabatic regime, such that Hawking evaporation sets in and competes with further accretion. We analyze the circumstances under which this evaporation can actually prevent the spacetime from reaching criticality. Secondly, in a configuration where the critical limit is surpassed, we will consider the impact of backreaction in the ultramassive Kottler region. We will look at the impact of the RSET on the generation of the future spacelike singularity, including the possible presence of a Hawking flux generated during a submassive phase. 

\subsection{Near criticality: accretion vs evaporation}
\label{Subsec:NearCriticality}
Let us consider a spacetime which begins its evolution analogously to the case outlined in subsection~\ref{Subsec:HawkingRadTwoShells}. An initially sub-massive black hole of $m_1$ is formed through gravitational collapse and settles into a (quasi-)stationary configuration. Then, the initial conditions at past infinity are set up such that subsequent accretion of a mass $m_2$ should, classically, make the transition to an ultramassive configuration, i.e. $m_\infty=m_1+m_2>m_\text{crit}$. The question then arises as to how the Hawking evaporation during the stationary sub-massive phase of the evolution can affect the ultimate fate of the spacetime.

For simplicity, we will again consider that $m_1$ and $m_2$ are in the form of ingoing null shells located at advanced times $v_1$ and $v_2$. If the accreted mass $m_2$ by itself is above the critical mass, $m_2>m_\text{crit}$, then the ultramassive limit is surpassed everywhere for $v>v_2$, making the future of the spacetime beyond this surface that of a cosmology collapsing into a singularity. The same occurs if $m_2<m_\text{crit}$ but the stationary period between the collapse of $m_1$ and $m_2$ is not long enough for the mass evaporated through the Hawking effect $m_\text{H}$ to bring the final mass below criticality, i.e. $m_1+m_2-m_\text{H}>m_\text{crit}$.

The construction with the most interesting outcome is one where the above inequality is reversed, i.e. one in which the Hawking evaporation tips the scale back down to a sub-massive configuration. Let us examine this case more closely.

As we have shown above, after the collapse of $m_1$, the RSET will have a negative ingoing flux component on the black hole horizon, which will drive its evaporation. In general, to have a full perturbative solution of the semiclassical equations, we would need to expand the conformal factor up to third order in the near horizon limit (meaning that we should expand Eq.~\eqref{Eq:Exp_Peeling} to third order around the horizon), see~\cite{Barcelo2020} for a careful analysis and discussion. However, to track the evolution of the size of the horizon, it is sufficient to consider only the ingoing flux $T_{vv}$ component, since it directly dictates the rate of evaporation in a quasi-stationary regime. The other components give rise to deformations of the space around the horizon, which we can ignore in an analysis of the causal structure.

\begin{figure}
    \centering
    \includegraphics[width=0.6\linewidth]{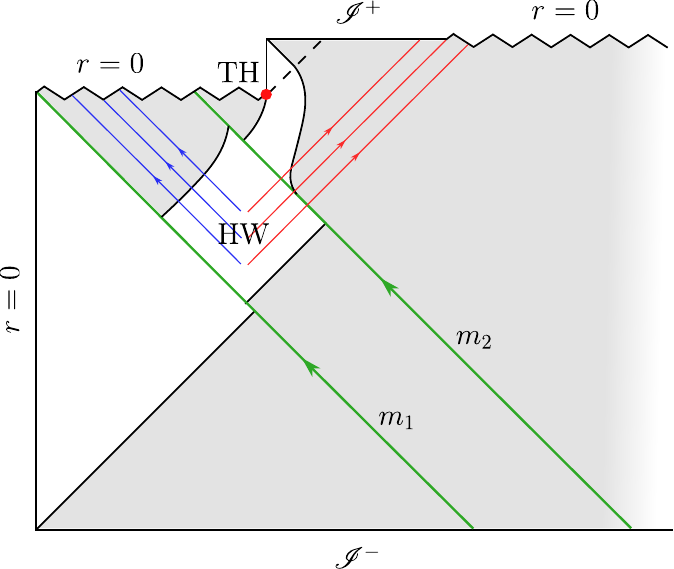}
    \caption{Black hole set to become ultramassive from initial conditions at $\mathscr{I}^-$, but which  evaporates semiclassically before the critical mass is reached. To the right the diagram exhibits the causality of an ultramassive Kottler spacetime, as this is part of the initial conditions. To the left, the black hole loses enough mass between shells $m_1$ and $m_2$ such that in the vicinity of the black hole the spacetime is not ultramassive. In the future region, the transition from sub-massive to ultramassive occurs along a line of outgoing Hawking radiation where the mass becomes critical. The shaded regions on the top left mark the black hole trapped region, and on the right they mark the cosmological region. The exterior of the black hole where Hawking radiation is first generated is marked with ``HW", and the ``thunderbolt" event at the end of the evaporation is indicated with ``TH" (the patch of spacetime in the causal future of this event is undetermined though in the diagram it is filled in with empty de Sitter spacetime).}
    \label{Fig:Evap}
\end{figure}

From Eq.~(38) of Ref.~\cite{Barcelo2020} we see that given the flux computed above, the perturbative change to the metric can be encoded in an expansion
\begin{align}
    f(v,r) \to f(v,r) + \delta f(v,r),
\end{align}
where $f(v,r)$ is the original function and $\delta f (v,r)$ is a perturbation sourced by the RSET. At leading order around the initial radial position of the black hole horizon $r_-$, this perturbation has the solution
\begin{equation}
    \delta f (v,r) = \frac{l_p^2\kappa_-^2}{24 \pi r_-} \left( v - v_{1} \right)+\order{r-r_-}+\order{(v-v_1)^2},
\end{equation}
where $v_1$ is the time at which the shell forming the black hole is located, $\kappa_-$ is the initial surface gravity of the horizon, and $l_p$ is the Planck length. This translates into a change of the black hole horizon position
\begin{equation}
    \delta r_-\simeq-\frac{\delta f|_{r=r_-}}{2\kappa_-}=-\frac{l_p^2\kappa_-}{48\pi r_-}(v-v_1)+\cdots
\end{equation}
Similarly, the positive outgoing flux $T_{u_>u_>}$~\eqref{tuu_h}, which eventually enters the cosmological region, induces a backreaction on the position of the (would-be) future cosmological horizon,
\begin{equation}
    \delta r_+\simeq\frac{\delta f|_{r=r_+}}{2\kappa_+}=\frac{l_p^2\kappa_-^2}{48\pi\kappa_+r_+}(u_{>}-u_0)+\cdots,
\end{equation}
where $\kappa_+$ is the surface gravity of the cosmological horizon (after the first shell), $r_+$ the radial position of this horizon, and $u_0$ some initial time at which we assume backreaction to begin. This backreaction can push the (would-be) future cosmological horizon far out enough that when the second mass $m_2$ is added, part of the configuration where backreaction has taken place (some region $u_{>} > u_{\text{crit}}$) can end up sub-critical. This possible scenario is depicted in Fig.~\ref{Fig:Evap}. 

In principle, one could include higher order corrections by performing an adiabatic expansion in $m(v)$ at the black hole horizon and $m(u)$ at the cosmological horizon, determining the corrections to the incoming flux at next orders, as it is done in Sec. 4 of~\cite{Barcelo2020}, but this would not qualitatively change the causal picture presented here. To definitively and precisely conclude under which conditions semiclassical effects can prevent the MTS of the spacetime from reaching the critical limit and forming an ultramassive spacetime (see Fig.~\ref{Fig:Evap}), a full numerical study would be required, akin to the one performed in~\cite{Boyanov2025} to unveil the fate of the inner horizon evaporation in asymptotically flat spaces.

\subsection{Evolution inside the ultramassive region}
\label{Subsec:TrappedRegion}
Let us now consider a configuration which classically surpasses the ultramassive limit, and focus on the semiclassical evolution near the future singularity $r=0$. In particular, this refers to the entire future end of the spacetime in cases in which evaporation does not bring part of the geometry below criticality or, in cases in which it does, to the region where the outgoing Hawking radiation is still insufficient to do so and reaches $r=0$ (see top-right of Fig.~\ref{Fig:Evap}). We focus on understanding whether semiclassical effects can alter this part of the causal structure as well, e.g., by driving the evolution of the geometry in the direction of avoiding the singularity. For that purpose we look at both the local terms of the RSET, computed directly on top of a Kottler geometry with $m > m_{\text{crit}}$, as well as the Hawking flux derived in Subsec.~\ref{Subsec:HawkingRadTwoShells}, which can appear from a transitory sub-massive regime.

In general, semiclassical backreaction can be calculated perturbatively by considering generic perturbations around the Kottler metric, and using the RSET as a source in the perturbed Einstein equations. However, for the purpose of understanding if semiclassical effects tend to prevent the formation of the future singularity, we can do something simpler. We can consider the Raychaudhuri equation~\cite{Wald1984} for the congruence of null geodesics with tangent vectors $V^{\mu}_{\pm}$ satisfying $g_{\mu \nu}V_{+}^{\mu} V_{-}^{\nu} = 1$, parametrized by the affine parameter $\lambda$:
\begin{align}
    \frac{\dd }{\dd \lambda} \Theta = - \frac{1}{2} \Theta^2 - \hat{\sigma}_{ab} \hat{\sigma}^{ab}  + \hat{\omega}_{ab} \hat{\omega}^{ab} - T_{ab} V_{\pm}^a V_{\pm}^b, 
\end{align}
where $\Theta$ is the expansion of the geodesics, $\hat{\sigma}_{ab}$ is the shear, and $\hat{\omega}_{ab}$ is the vorticity  (we recall that they are projected onto the 2D space orthogonal to the vectors $V^{\mu}_{\pm}$). We have also replaced the Ricci tensor with the stress-energy tensor since the vector is null and the contribution from its trace and the cosmological constant drops. If we consider a congruence such that $\hat{\omega}_{ab} = 0$ the fate of the geometry, i.e., whether $\Theta \to - \infty$ in finite affine parameter, will be completely determined by the behavior of $T_{ab} V_{\pm}^a V_{\pm}^b$.

For the RSET, from the results of Subsec.~\ref{Subsec:HawkingRadTwoShells} it is evident that we can construct an Unruh-like state which accounts for the Hawking radiation produced by simply choosing the $F$ and $G$ functions (and the constant $\lambda$) used for de Sitter space (see Subsec.~\ref{Subsec:RSET_Calc}), with the addition of a term $\kappa_{-}^2/(48 \pi)$ in $F$, corresponding to the non-local contribution associated with the incoming Hawking radiation. While this choice would not lead to a regular RSET on all the horizons of a maximally extended Kottler geometry, it does reproduce the RSET well in the region of interest.

We consider the vectors $V^\mu_{\pm} = \pm f(T) \delta^{\mu}_{\ T} + \delta^{\mu}_{\ X}$, for which the vorticity vanishes, $\hat{\omega}_{ab} = 0$, and so does the contribution from the classical energy-momentum tensor. At first order in backreaction, we can take the RSET and the vectors $V^\mu_{\pm}$ obtained with the Kottler background geometry. Computing the contraction explicitly we find:
\begin{align}
    & T_{ab} V^a_{-} V^b_{-} = -  \frac{l_p^2m}{8 \pi^2  \ell^2 T^3}+\frac{l_p^2m^2}{16 \pi^2  T^6}+\frac{l_p^2m}{24 \pi^2  T^5}, \\
    & T_{ab} V^a_{+} V^b_{+} = \frac{l_p^2\kappa_-^2}{48 \pi^2 T^2} - \frac{l_p^2m}{8 \pi^2  \ell^2 T^3}+\frac{l_p^2m^2}{16 \pi^2 T^6}+\frac{l_p^2m}{24 \pi^2  T^5}.
\end{align}
From these expressions we can read the behavior of the RSET in these null directions. We first note that at past infinity \mbox{$T \to - \infty$} the contribution from the RSET decays to zero rapidly (noting also that to recover the RSET in this limit for the physical vacuum state of interest we should take $\kappa_- = 0$). Toward the future singularity $T \to 0^{-}$, both expressions diverge, with the strongest terms ($\sim T^{-6}$) being positive, which would contribute to the evolution by accelerating the focusing of the null geodesics and the approach to $\Theta \to - \infty$ in finite affine parameter, making the collapse faster.

As for the term with $\kappa_-$, corresponding to the flux of outgoing Hawking radiation, it also has a positive contribution, and thus also potentiates the divergence. While it is not the dominant contribution toward $T\to 0$, it would be relevant at intermediate values of $T$.

In fact, a very important point to note is that the Polyakov approximation for the RSET used here breaks down close to the $T\to 0$ limit. On the other hand, this approximation works well to capture the Hawking evaporation effect. Therefore, the term with $\kappa_-$ can be considered the most robust part of the result, and likely would persist in much the same form if a more precise computation of the RSET in $3+1$ dimensions is performed. The fact that this term can dominate the RSET at intermediate times, and that it is positive, implies that semiclassical effects would, at least transitorily, accelerate the collapse. Determining whether the dominant terms at $T\to 0$ from a $3+1$ computation of the RSET would also potentiate the collapse falls beyond the scope of the present work.

However, we should remark that even though the Polyakov approximation is beyond its regime of applicability in this case, it still exhibits a qualitative behavior close to what is to be expected based on the recent analysis reported in~\cite{Gao2023}. There, a slightly different situation is considered. Semiclassical effects (including the backreaction) are studied on top of de Sitter spacetime regarded as an FLRW cosmological spacetime (in the sense of~\cite{Garcia-Moreno2026}) with compact $\mathbb{S}^3$ slices that are initially contracting and such that in the absence of semiclassical effects, would bounce back and infinitely expand. Notice that it corresponds to the $m = 0$ case of the Kottler geometry. In this case, the authors find that if enough particles are produced, their backreaction makes the spacetime develop a Big Crunch-like singularity in the future. In fact, the metric looks precisely like the Kottler metric with $m > m_{\text{crit}}$. Hence the expectation is that once semiclassical effects are included on top of one of this, it is impossible to prevent the development of a singularity, as the particle production tends to make the geometry collapse. In addition to this effect that is qualitatively captured by the Polyakov approximation, a more sophisticated approximation of the RSET (notice that within the Polyakov approximation, the angular components identically vanish) is expected also to dilute the anisotropies and result in a configuration that approaches an isotropic state~\cite{Hu1978}. 

\section{Conclusions}
\label{Sec:Conclusions}

In this paper we have presented the first study of semiclassical effects on the recently introduced ultramassive spacetimes. We started by reviewing the construction of some of the ultramassive spacetimes in spherical symmetry presented in~\cite{Senovilla2022}, which we then used as background geometries for our analysis. The first difficulty we faced was in constructing an adequate vacuum state for free quantum fields propagating in these spacetimes. Particularly, the absence of a privileged vacuum state for a field propagating on de Sitter (this being the limit in the asymptotic past of the geometries in question) makes it difficult to define an adequate ``in" state. Following the logic from asymptotically flat spacetimes, we chose the vacuum state, along with the renormalization constants of the RSET, for which the RSET vanishes in de Sitter spacetime, and we used this choice at $\mathscr{I}^{-}$ in the geometries which represent gravitational collapse of initially diffuse matter. For the RSET, we used the spherical Polyakov approximation for a massless scalar field, for which the above conditions lead to a unique state and renormalization constant.

For the geometry, we considered a Kottler-Vaidya spacetime, where a black hole forms at some finite time, with subsequent accretion possibly driving it past criticality and into an ultramassive configuration. We calculated the values of the RSET in the region outside the (initially) submassive black hole, finding the fluxes which are typically identified with the production of Hawking radiation: a negative ingoing flux falling into the black hole horizon, and a positive outgoing flux which carries the ``evaporated" mass outward. It is important to note that the computation of the RSET here is the most natural method for determining that Hawking radiation is produced, since the standard computation of Bogoliubov coefficients between ``in" and ``out" states requires matching the natural modes in the asymptotic past and future null regions, the latter possibly being absent here. Much like in asymptotically flat spacetime, the presence of these fluxes is a robust result not only during stationary periods of the evolution, but also during sufficiently slow accretion, controlled by an adiabatic condition. In other words, a flux of Hawking radiation arises if there is a sufficiently long period of adiabatic accretion. In such a scenario, the computation of the RSET and the analysis of perturbative backreaction performed are straightforwardly generalizable.

Regarding backreaction, we firstly analyzed the perturbation to the evolution of the black hole and cosmological horizons caused by the fluxes associated with Hawking radiation. As may be expected, we found that (under the assumption of adiabatic evolution) the black hole horizon shrinks due to the negative ingoing flux, while the cosmological horizon expands due to the positive outgoing flux (which leaves a configuration with less mass in its wake). This tendency continues for as long as semiclassical backreaction dominates over any possible tenuous classical accretion. Interestingly, if classical initial conditions are set up such that a subsequent accretion drives the configuration past the critical limit, but evaporation goes on for long enough (i.e. the adiabatic state of the black hole continues for long enough) to remove enough mass so as to bring the asymptotic mass below this limit, the asymptotic future of the spacetime can be split into a region which recovers the $\mathscr{I}^+$ of de Sitter, and a region where the future singularity of the ultramassive configuration persists. Considering only backreaction from the flux terms of the RSET, these regions would be separated by the ray of outgoing Hawking radiation which brings the remaining mass under the critical limit. In a complete self-consistent solution, this transition may be more complex. 

We have also considered backreaction in the collapsing cosmological region where the future singularity of the ultramassive configuration is formed. While the Polyakov approximation is not guaranteed to be adequate in this region, the terms associated with the Hawking evaporation are at least expected to be the same in more precise computations of the RSET. In any case, both the local terms from the Polyakov approximation and the non-local (and more robust) Hawking flux terms seem to accelerate the collapse, which agrees qualitatively with the results presented in Ref.~\cite{Gao2023} where a similar analysis was presented with other motivations. 

We note that classically the critical case corresponds to the formation of a zero surface gravity black hole and hence would constitute a violation of the third law. This case is of course a fine tuned situation in which the mass needs to be precisely balanced with the cosmological constant. It is then not unreasonable to expect that semiclassical effects may always avoid the formation of this configuration. If one throws a matter content that is too heavy, the evaporation of the horizon will not be fast enough to avoid entering into the collapsing region and the end state will still be an ultramassive configuration. If one throws a matter content slightly heavier than the critical mass, the evaporation due to Hawking radiation may actually be fast enough to make the black hole horizon shrink too fast and simply evaporate.

Here we have considered the simplest approximation that one can conceive to compute the RSET, namely the Polyakov approximation and we have only considered the perturbative backreaction, without fully evolving the semiclassical equations in a self-consistent way. There are two natural extensions of this work that are worth exploring in the future. First of all, one can use more sophisticated approximations to compute the RSET, as for instance the $s$-wave approximation presented in Ref.~\cite{Fabbri2005}. The improvement of the $s$-wave approximation with respect to the Polyakov approximation is that it includes the backscattering of the modes due to the potential in the mode equation. This would be particularly interesting as it would be a controlled approximation within the fully trapped region, where it is not clear that the backscattering can be neglected. Second, it would be interesting to evolve self-consistently the semiclassical equations as done in Ref.~\cite{Boyanov2025}. In particular, it will be quite interesting to consider initial data that are close to the critical mass and check whether the possibility discussed above is true, namely, whether semiclassical effects always prevent the formation of a zero surface gravity and, hence, avoids the violation of the third law.

%

\section*{Acknowledgements}
\addcontentsline{toc}{section}{Acknowledgements}

We thank José M. M. Senovilla for raising the questions that originally motivated this work, for very useful discussions, and for carefully reading the manuscript. We also thank Aitor Vicente-Cano, Carlos Barcel\'o, Robin Croft, Alexandru Dima, Paolo Pani, Silvia Pla, Jos\'e Polo-G\'omez and Alfredo Stanzione for helpful discussions. 
G. Garc\'ia-Moreno acknowledges support from the MUR FIS2 Advanced Grant ET-NOW (CUP: B53C25001080001) and the INFN TEONGRAV initiative. 
A.G.M.-C. acknowledges support from the Xunta de Galicia  through Grant No.~ED431B-2024/42 and the postdoctoral fellowship Grant No.~ED481B-2025/059.
V.B. acknowledges support from the European Union’s H2020 ERC Advanced Grant “Black holes: gravitational engines of discovery” grant agreement no. Gravitas–101052587.
Financial support was also provided by the Spanish Government Grants PID2023-149018NB-C43 and PID2023-149018NB-C44 (funded by MICIU/AEI/10.13039/501100011033  and by ``ERDF/EU A way of making Europe'').

\newpage

\appendix

%
\section{Quantum field theory in de Sitter spacetime}
\label{App:deSitter}
Consider the five-dimensional Minkowski spacetime covered with the cartesian coordinates $Y^{M}$, where $M = 0,1,\ldots,4$ in terms of which the line element is:
\begin{align}
   \dd s^2 =  \eta_{MN} Y^M Y^N = - \left( \dd Y^0 \right)^2 + \sum_{I =1}^{4} \left( \dd Y^I \right)^2.
   \label{Eq:Flat5D}
\end{align}
We define the four-dimensional de Sitter spacetime as the four-dimensional embedded manifold described by:
\begin{align}
    - \left( Y^0 \right)^2 + \sum_{I =1}^{4} \left( Y^I \right)^2 = \ell^2,
\end{align}
with the metric that inherits from Eq.~\eqref{Eq:Flat5D}. It is a solution to Einstein equations with a positive cosmological constant which is related to $\ell$ by $ \Lambda = 3 / \ell^2$. 

We focus here on the past patch that connects with $\mathscr{I}^{-}$ that is depicted in Fig.~\ref{Fig:deSitter}. For that purpose, we can introduce coordinates $(T,X, \Omega)$ (where $\Omega$ represents coordinates in the unit two-sphere) by simply taking:
\begin{align}
    & Y^0 = \sqrt{T^2 - \ell^2} \cosh \left( \frac{X}{\ell} \right) , \\
    & Y^1 = \sqrt{T^2 - \ell^2 } \sinh \left( \frac{X}{\ell} \right) , \\
    & Y^I = T \hat{n}^I ( \Omega ), \qquad I = 2, 3, 4,
\end{align}
where $\hat{n}^I ( \Omega )$ is the unit normal vector to the two-sphere. We then recover the line element~\eqref{Eq:deSitterAsympt}:
\begin{align}
    \dd s^2 = - \frac{\dd T^2}{\abs{f(T)}} + f(T) \dd X^2 + T^2 \dd \Omega^2,
\end{align}
where $f(T) = T^2/\ell^2 -1$ and the coordinates span $ - \infty < T < - \ell$ and $ - \infty< X < \infty$\footnote{Alternatively we could take $ \ell < T < \infty$ but that case corresponds to the future region in which we are not interested here.}. From now on, we focus on the two-dimensional sector of the metric as the radius of the spheres is always well-behaved in our changes of coordinates. In the main text, we have introduced coordinates $(\tau, X)$ defined in Eq.~\eqref{Eq:taudesitter} from which we can change to $(u,v)$ introduced below Eq.~\eqref{Eq:taumetricdesitter} which are valued on the whole real line but with the constraint $u+v < 0$, so that the conformal factor does not blow up.  

Notice that here we are singling out a family of observers, namely those in orbits of $\partial_T$ which are those with respect to which the quantization is adapted. In fact, the isometry group of de Sitter spacetime is $SO(1,4)$\footnote{This is specially clear from the isometric embedding of the hyperboloid in Minkowski spacetime described above.} and there is no unique way of decomposing the metric as a warped product like the one introduced in Eq.~\eqref{Eq:2D_metric}. In other words, there is no intrinsic foliation in spheres of some of the spatial coordinates since there is no unique $SO(3) \subset SO(1,4)$. In that sense, the Polyakov approximation is explicitly breaking the symmetries by singling out a family of observers. However, this will not be relevant for our other considerations since in the dynamical cases the symmetry group is not as big and such intrinsic foliation in spheres does exist.

From the embedding perspective here described, it is clear that the geometry is smooth across the cosmological horizons. To make it explicit, we notice that we can make the change of coordinates to $(U,V)$ defined as:
\begin{align}
    &  U = - e^{-u/\ell}, \qquad d U = \frac{du}{\ell} e^{-u/\ell}, \\
    &  V = - e^{-v/\ell}, \qquad d V = \frac{dv}{\ell} e^{-v/\ell},
\end{align}
in terms of which the 2D sector of the metric can be expressed as
\begin{align}
    \dd s^2 = - \frac{4 \ell^2}{(1-UV)^2} \dd U \dd V.
\end{align}
The range of original coordinates is now $U \in (-\infty, 0) $ and $V \in (-\infty, 0)$ with $UV < 1$. However, now we can extend the metric beyond $ U = 0$ and $V = 0$, which is where the horizons are located. 

If we take the expressions for the RSET in Eqs.~\eqref{Eq:Tuudesitter}--\eqref{Eq:Tvvdesitter} we can compute it in terms of $C(u,v)= - \sinh^{-2} \left( \frac{u+v}{2 \ell} \right)$ and change it to the $(U,V)$ coordinates:
\begin{align}
    & T_{UU} = \frac{\ell^2}{U^2} \left( - \frac{1}{48 \pi \ell^2} + F\left(  u(U) \right) \right)  , \\
    & T_{UV} =  \frac{1}{(1-UV)^2} \left(\frac{1}{12 \pi \ell^2}- {2 \lambda \ell^2}\right),  \\
    & T_{VV} = \frac{\ell^2}{V^2} \left( - \frac{1}{48 \pi \ell^2} + G\left(  v(V) \right) \right).
\end{align}
The flux components $T_{UU}$ and $T_{VV}$ are singular at the horizon for any choice of $F$ and $G$ except for
\begin{align}
    F(x) = G(x) = \frac{1}{48 \pi \ell^2}.
\end{align}
In fact, it is possible to compute the RSET directly using the $(U,V)$ coordinates, for which $C(U,V) = 4 \ell^2 (1-UV)^{-2}$, and we find
\begin{align}
    & T_{UU} = \Tilde{F}(U), \\
    & T_{UV} = \frac{1}{12 \pi \ell^2} \frac{1}{(1-UV)^2} - \frac{2 \lambda \ell^2}{(1-UV)^2},  \\
    & T_{VV} = \Tilde{G} (V).
\end{align}
We see that the vacuum we have chosen before, here simply corresponds to the choice $\Tilde{F}(U) = \Tilde{G} (V) = 0$ and $\lambda = 1/24 \pi \ell^2$, i.e., the coordinates $(U,V)$ are precisely adapted to the vacuum choice that we are making, as discussed in Subsec.~\ref{Subsec:RSET_generalities}.

\section{Hawking radiation in Vaidya spacetime}
\label{App:Flat_Collapse}
Consider the Vaidya line element on top of flat spacetime which formally corresponds to the limit $\ell \to \infty $ of the Vaidya-de Sitter line-element presented in Eq.~\eqref{Eq:VaidyadeSitter}, i.e., we have
\begin{align}
    \dd s^2 = - \left( 1 - \frac{2m(v)}{r} \right) \dd v^2 + 2 \dd v \dd r + r^2 \dd \Omega^2. 
\end{align}
In general, to fully determine the RSET within the Polyakov approximation everywhere in the spacetime we would need to explicitly find the change of coordinates to double null coordinates $(v,r) \to (v,u)$ such that the metric is given by:
\begin{align}
    \dd s^2 = - C(u,v) \dd u \dd v + r^2(u,v) \dd \Omega_2^2,
\end{align}
where $C(u,v) = 2 \partial_u r (u,v)$. Doing so in general is an impossible analytic task, as it implies solving the equation:
\begin{align}
    \frac{\partial r}{\partial v} = \frac{1}{2}  \left( 1 -  \frac{2m(v)}{r} \right),
    \label{Eq:evolution}
\end{align}
for a given $m(v)$ profile. It is possible to solve this equation when $m (v) = m$, i.e, for constant masses\footnote{There is another case that can be solved in which the mass function is $m(v) = \alpha v$. In this case, the geodesic equation can also be solved and an implicit relation between $r$ and $v$ can be found~\cite{Waugh1986}.}, we get the well-known solution:
\begin{align}
    r_{\ast} (r) = r + 2m \log\left( \frac{r}{2m} -1 \right) = \frac{v}{2} + I(u),
\end{align}
up to an arbitrary function of $u$ which labels different lightrays. This specifies (implicitly in the case of finite constant $m$) the change of coordinates $r(v,u)$. It is convenient to chose $I(u) = - u/2$ so that both coordinates are defined symmetrically. 

\begin{figure}
    \centering
    \includegraphics[width=0.35\linewidth]{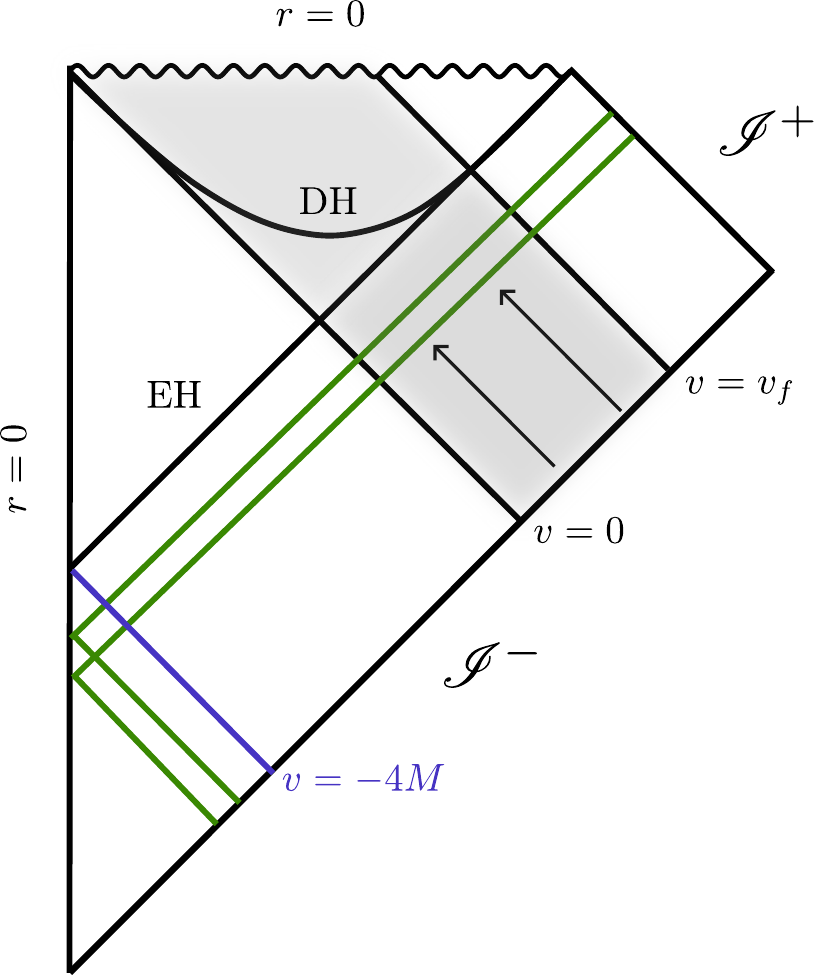}
    \caption{We represent here the Penrose diagram for the de Vaidya collapse in flat spacetime. We have shaded in grey the region where $m' \neq 0$. The metric below   is flat and the metric above   is a patch of Schwarzschild spacetime. EH represents the event horizon $r = 2 m_{\infty}$, and DH the trapped region which increases its radius from $0$ to $2m_{\infty}$, joining the event horizon asymptotically. The blue lightray represents the generator of the horizon and the green curves are two ingoing lightrays that, although initially separated a distance in affine parameter, they end up exponentially separated at the asymptotic time.}
    \label{Fig:Vaidya_Flat}
\end{figure}

In general, we are interested in the case in which there is a finite window of time $(v_0,v_f)$ for which $m'(v)\neq 0$, i.e., the black hole is formed within a finite window of time. For the sake of simplicity, from now on we take $v_0 = 0$. We assume that condition~\eqref{Eq:BoundVaidya} holds so that there are no naked singularities~\cite{Kuroda1984}. If we start with $m(v) = 0$ for $v<0$, the natural vacuum choice is the in vacuum that arises from the change of coordinates to flat spacetime double null coordinates:
\begin{align}
    u_{<} = v - 2 r, 
\end{align}
for which the $r(u_{<},v) = (v-u_{<})/2$ and $C(u_{<},v) = 1$, valid for $v < 0$. At $v = 0$, this gives the data required to specify the integration function for the change to double null coordinates, i.e., we must solve Eq.~\eqref{Eq:evolution} with the initial condition $r(u_{<},0) = - u_{<}/2$. The Penrose diagram is represented in Fig.~\ref{Fig:Vaidya_Flat}.

\subsection{Thin-shell collapse}
\label{Subsec:ThinShell}
As an illustrative example, we consider the case in which $m(v) = m_{\infty} \Theta (v)$, i.e., the null radiation is supported only on the null ray $v=0$. It can be understood as a distributional limit in which $v_f \to 0$. In this case, choosing the $u_{>}$ coordinate (for $v >0$) symmetrically with respect to $v$ leads to:
\begin{align}
    r_{\ast} (u_{>}, v) = r (u_{>},v) + 2 m_{\infty} \log \left( \frac{r(u_{>},v)}{2m_{\infty}} \right) = \frac{1}{2} (v-u_{>}),
    \label{Eq:TortoiseOut}
\end{align}
where we have introduced the Regge-Wheeler coordinate $r_{\ast}$. In terms of this coordinate, the metric is given by $C(u_{>},v) = 1 - \frac{2m_{\infty}}{r(u_{>},v)}$, with $r(u_{>},v)$ defined implicitly in Eq.~\eqref{Eq:TortoiseOut}. Continuity of the metric requires:
\begin{align}
    r \left( u_{<}, 0 \right) = r \left( u_{>}, 0 \right).
    \label{Eq:MatchingVaidya}
\end{align}
Thus, from Eq.~\eqref{Eq:TortoiseOut} we find
\begin{align}
    r (u_{>}, 0) + 2 m_{\infty} \log \left( \frac{r (u_{>}, 0)}{2m_{\infty}} \right) = - \frac{u_{>}}{2},
\end{align}
and using Eq.~\eqref{Eq:MatchingVaidya} to substitute $r \left( u_{>}, 0 \right)$ by $r \left( u_{<}, 0 \right)$, we get 
\begin{align}
    r (u_{>}, 0) + 2 m_{\infty} \log \left( \frac{r (u_{>}, 0)}{2m_{\infty}} \right) = - \frac{u_{<}}{2} + 2 m_{\infty} \log \left( - \frac{u_{<}}{4m_{\infty}} \right) = - \frac{u_{>}}{2},
\end{align}
which means that we have the following matching:
\begin{align}
    u_{>} = u_{<} - 4m_{\infty} \log \left( - \frac{u_{<}}{4m_{\infty}} \right).
    \label{Eq:ThinShellinout}
\end{align}
Applying the chain rule to express everything in terms of $u_{<}$ we get the following line element:
\begin{align}
    \dd s^2 = - C(u_{<},v) \dd u_{<} \dd v + r^2(u_{<},v) \dd \Omega_2^2,
\end{align}
where
\begin{align}
    C(u_{<},v) = \begin{cases} 
      1, & v < 0, \\
      \left(1 - \frac{2m_{\infty}}{r(u_{<},v)}\right) \frac{u_{<}}{u_{<}+4m_{\infty}},  & v > 0, 
   \end{cases}
\end{align}
and $r(u_{<},v)$ is determined (implicitly for $v>0$) by
\begin{align}
    \begin{cases} 
      r(u_{<},v) =  \frac{v-u_{<}}{2}, & v \leq 0, \\
      r(u_{<},v) + 2m_{\infty} \log \left( \frac{r(u_{<},v)}{2m_{\infty}} - 1 \right) = \frac{v-u_{<}}{2},  & v > 0. 
   \end{cases}
\end{align}
We are interested in the asymptotic region $u_{>} \to \infty$, for fixed $v$. It corresponds to $r \to \infty$, which is tantamount to $u_{<} \to - 4m_{\infty}$, hence from the conformal factor we can focus only on the term arising from $du_{>}/du_{<}$ which gives the following only nonvanishing component as $u_{<} \to -4 m_{\infty}$:
\begin{align}
    T_{u_{<}u_{<}} = \frac{1}{48 \pi \left(u +4m_{\infty} \right)^2} + \order{\frac{1}{(u+4m_{\infty})}},
\end{align}
which after changing to $u_{>}$ coordinate reads
\begin{align}
    T_{u_{>}u_{>}} = \frac{1}{768 \pi m_{\infty}^2 },
\end{align}
i.e., the flux of Hawking radiation associated with blackbody radiation for temperature $1/(8 \pi m_{\infty})$.

\bibliographystyle{JHEP}
\bibliography{ultramassive_biblio}

\end{document}